\documentclass[showpacs,aps,prl,reprint,
superscriptaddress,longbibliography,
]{revtex4-1}
\usepackage[colorlinks=true, pdfstartview=FitV,
bookmarks=true, bookmarksnumbered=true, breaklinks]{hyperref}
\usepackage[dvipdfmx]{graphicx}
\usepackage{amsmath,amssymb,bm,color,longtable,mathrsfs,slashed,comment,tikz}

\definecolor{darkviolet}{rgb}{0.58, 0.0, 0.83}
\definecolor{electricultramarine}{rgb}{0.25, 0.0, 1.0}
\definecolor{brightpink}{rgb}{1.0, 0.0, 0.5}
\hypersetup{linkcolor=brightpink, citecolor=darkviolet, urlcolor=electricultramarine}

\definecolor{lime}{HTML}{A6CE39}
\DeclareRobustCommand{\orcidicon}{
	\hspace{-3mm}
	\begin{tikzpicture}
	\draw[lime, fill=lime] (0,0) 
	circle [radius=0.16] 
	node[white] {{\fontfamily{qag}\selectfont \tiny ID}};
	\draw[white, fill=white] (-0.0625,0.095) 
	circle [radius=0.007];
	\end{tikzpicture}
	\hspace{-3mm}
}

\foreach \x in {A, ..., Z}{\expandafter\xdef\csname orcid\x\endcsname{\noexpand\href{https://orcid.org/\csname orcidauthor\x\endcsname}
			{\noexpand\orcidicon}}
}

\begin{document}

\title{Gravitational form factors of pion from top-down holographic QCD}

\author{Daisuke~Fujii\orcidB{}}
\email[]{daisuke@rcnp.osaka-u.ac.jp}
\affiliation{Advanced Science Research Center, Japan Atomic Energy Agency (JAEA), Tokai, 319-1195, Japan}
\affiliation{Research Center for Nuclear Physics, Osaka University, Ibaraki 567-0048, Japan}

\author{Akihiro~Iwanaka\orcidC{}}
\email[]{iwanaka@rcnp.osaka-u.ac.jp}
\affiliation{Research Center for Nuclear Physics, Osaka University, Ibaraki 567-0048, Japan}

\author{Mitsuru Tanaka\orcidA{}}
\email[]{tanaka@hken.phys.nagoya-u.ac.jp}
\affiliation{Department of Physics, Nagoya University, Nagoya 464-8602, Japan}

\begin{abstract}

The gravitational form factors (GFFs) of pions are calculated from a top-down holographic quantum chromodynamics (QCD) approach with momentum transfer dependence for the first time. It is important because the GFFs of hadrons have information on the internal stress distribution that may provide insight into the mechanism of how QCD forms hadrons. The forward limit value of this GFFs, i.e. the D-term, was also obtained. Furthermore, in this approach, we observe the so-called glueball dominance, in which pions have gravitational interactions via infinite glueball spectra. 

\end{abstract}

\maketitle

{\it Introduction.}---
How forces act inside hadrons to confine quarks and gluons forming hadrons has not been solved. 
Addressing this question is inseparable from developing an insight into the non-perturbative properties of quantum chromodynamics (QCD), such as confinement and spontaneous chiral symmetry breaking, which is a crucial problem in modern physics. 

Recently, the stress inside hadrons has attracted attention (see a recent review \cite{Polyakov:2018zvc,Burkert:2023wzr}) due to the fact that experimentally it has been measured. The stress distribution inside nucleons, extracted from the matrix elements of the nucleon energy-momentum tensor (EMT), is obtained from deeply virtual Compton scattering~\cite{Diehl:2003ny,Belitsky:2005qn,Mezrag:2022pqk}. 
The form factors that characterize the matrix elements are called gravitational form factors (GFFs), which are calculated theoretically from their coupling with gravity. 

The forces inside nucleons were analyzed experimentally in Ref.~\cite{Burkert:2018bqq} for the quark part in 2018, and in Ref.~\cite{Duran:2022xag} for the gluon part in 2022. 
Experimental results show that around the center of the nucleon, the force pushes the quarks outward, while around the surface it reverses into a confining force, indicating an extremely anisotropic behavior. As a result, the pressure near the center of the nucleon may reach the highest value currently observed in nature. The pion GFFs have also been analyzed experimentally in Ref.~\cite{Belle:2012wwz,Belle:2015oin,Kumano:2017lhr}. The GFFs of nucleons~\cite{LHPC:2007blg,Bali:2016wqg,Shanahan:2018pib,Alexandrou:2018xnp,Alexandrou:2019ali,Alexandrou:2020sml,Pefkou:2021fni,Hackett:2023rif,HadStruc:2024rix} and pions~\cite{Brommel:2005ee,Brommel:2007zz,Yang:2014xsa,Shanahan:2018pib,Pefkou:2021fni,Loffler:2021afv,Hackett:2023nkr} have also been measured by lattice QCD analysis, a first-principles numerical simulation of QCD. 

In this letter, we are the first to calculate the momentum transfer dependence of the pion GFFs from a top-down holographic QCD approach. We use the Sakai-Sugimoto model, a top-down holographic QCD model based on the D4-D8 brane construction~\cite{Sakai:2004cn,Sakai:2005yt}. With this approach, the spectroscopy~\cite{Constable:1999gb,Brower:2000rp,Sakai:2004cn,Hata:2007mb,Hashimoto:2009hj,Hashimoto:2009st,Imoto:2010ef,Ishii:2010ib,Liu:2017xzo,Liu:2017frj,Hashimoto:2019wmg,Liu:2019yye,Nakas:2020hyo,Hayashi:2020ipd,Fujii:2020jre,Suganuma:2020jng,Liu:2021tpq,Liu:2022urb}, decay widths and electromagnetic form factors~\cite{Sakai:2005yt,Hashimoto:2007ze,Hashimoto:2008zw,Hata:2008xc,Kim:2008pw,Grigoryan:2009pp,BallonBayona:2009ar,Bayona:2009pk,BallonBayona:2010ae,Cherman:2011ve,Bayona:2011xj,Harada:2011ur,Brunner:2015oqa,Li:2015oza,Druks:2018hif,Fujii:2021tsw,Liu:2021ixf,Liu:2021efc,Iwanaka:2022uje,Fujii:2022yqh,Fujii:2023ajs,Bigazzi:2023odl,Hechenberger:2023ljn,Ramalho:2023hqd,Castellani:2024dxr,Hechenberger:2024piy} of mesons, baryons, and glueballs have been investigated with strong coupling and large $N_c$ limit. Recently, using the same approach, nucleon GFFs have been formulated and D-terms have been estimated~\cite{Fujita:2022jus}. The GFFs of pions~\cite{Abidin:2008hn} and other hadrons~\cite{Abidin:2008ku,Abidin:2009hr,Mamo:2019mka,Chakrabarti:2020kdc,Mamo:2021krl,Mamo:2022eui,Allahverdiyeva:2023fhn} have also been calculated using bottom-up holographic QCD, which has no direct connection with QCD through string theory. 

It is interesting that stress distributions and GFFs of pions\cite{Polyakov:1999gs,Broniowski:2008hx,Frederico:2009fk,Masjuan:2012sk,Son:2014sna,Fanelli:2016aqc,Hudson:2017xug,Freese:2019bhb,Krutov:2020ewr,Shuryak:2020ktq,deTeramond:2021lxc,Raya:2021zrz,Tong:2021ctu,Krutov:2022zgg,Xu:2023izo,Li:2023izn,Broniowski:2024oyk,Liu:2024jno,Liu:2024vkj}, which are the focus of this letter, show anomalous behavior. 
The GFFs are very sensitive to interactions in the system. The forward limit of the GFFs giving the stress distribution is called the D-term, which takes the value $-1$ for spin-zero particles in the free Klein-Gordon theory. On the other hand, even an infinitesimally small interaction causes a significant deviation from $-1$~\cite{Hudson:2017xug}. 
However, the D-term of the Nambu-Goldstone (NG) bosons in the strongly coupled system becomes $-1$ like the free particles, in the soft pion limit because of the restriction by chiral symmetry~\cite{Donoghue:1991qv,Polyakov:1999gs,Hudson:2017xug}. 

This fact may provide some insight into the role of chiral symmetry breaking in hadron formation since the D-term carries information about the forces inside hadrons. 
These insights are expected to be validated in more detail via analysis of hadron GFFs from future hadron experiments~\cite{Burkert:2018nvj,AbdulKhalek:2021gbh,Anderle:2021wcy,JeffersonLabHallA:2022pnx,CLAS:2022syx}. 

\vspace{\baselineskip}

{\it Formulation.}---
In this section, we provide the method to derive the GFFs of pions in the top-down holographic QCD~\cite{Sakai:2004cn,Sakai:2005yt}. 
The method for calculating the GFFs of nucleons with spin 1/2 has been formulated in Ref.~\cite{Fujita:2022jus} and will be applied in the following.

The matrix elements of EMT of a spin-0 particle are expressed as
\begin{align}
    &\big<\pi^a(p_2)|\hat{T}^{\mu\nu}(x)|\pi^b(p_1)\big> \notag \\
    &=\delta^{ab}\Big[2P^\mu P^\nu A(t)+\frac{1}{2}(\Delta^\mu\Delta^\nu-\eta^{\mu\nu}\Delta^2)D(t)\Big]e^{ix\cdot\Delta}, \label{GFFsspin0}
\end{align}
where $a,b$ are the indices for isospin and the kinematic variables are defined as $P^\mu=(p^\mu_1+p^\mu_2)/2$, $\Delta^\mu=p^\mu_2-p^\mu_1$ and $t=\Delta^2$ with Lorentz indices $\mu,\nu$~\cite{Kobzarev:1962wt,Pagels:1966zza,Polyakov:2018zvc}. 
In this paper, we employ the mostly plus metric $\eta^{\mu\nu}={\rm diag}(-1,1,1,1)$ and the normalization condition as $\big<\vec{p_2}|\vec{p_1}\big>=2E(2\pi)^3\delta^{(3)}(\vec{p_1}-\vec{p_2}), \ \ \ E=\sqrt{\vec{p_1}^2+m^2}$.

The method for calculating the matrix elements of the EMT by holographic QCD is as follows. 
Within the framework of holographic QCD, perturbations to the metric caused by matter fields in the bulk propagate to the boundary, allowing us to extract the expectation values of the EMT in the boundary theory from these perturbations. The top-down holographic QCD used in this study, the Sakai-Sugimoto model, consists of $N_f$ D8 branes embedded as a probe in black 4-brane background~\cite{Witten:1998zw}, a holographic dual of $N_c$ D4 branes. This background is a 10-dimensional curved spacetime of type IIA supergravity, derived from 11-dimensional supergravity related to M-theory on doubly Wick rotated $\rm AdS_7$ black hole $\times S^4$, 
\begin{align}
    ds^2=& \frac{r^2}{L^2}\big[f(r)d\tau^2-dx^2_0+dx_1^2+dx^2_2+dx^2_3+dx^2_{11}\big] \label{AdS7BH} \\ 
    &+\frac{L^2}{r^2}\frac{dr^2}{f(r)}+\frac{L^2}{4}d\Omega^2_4 , \notag \\
    f(r)=& 1-\frac{R^6}{r^6}, \ \ \ R = \frac{L^2M_{KK}}{3}, \ \ \ L^3=8\pi g_s N_c l^3_s, \ \ \ \notag \\
    R_{11}=&
    \frac{\lambda}{2\pi N_c M_{KK}}, \ \ \ \lambda N_c=\frac{L^6M_{KK}}{32\pi g_s l^5_s}=216\pi^3\kappa, \notag
\end{align}
where $M_{KK}$ is the Kaluza-Klein mass in the $\tau$ direction, $g_s$ string coupling, $l_s$ string length, $\lambda$ 't Hooft coupling, and $R_{11}$ the radius of the circle compactified in the $x_{11}$ direction.

The expectation values of EMT on the boundary, $\big<T_{\mu\nu}\big>$, can be extracted from the boundary values of the metric fluctuations in the bulk in the background spacetime \eqref{AdS7BH} propagated to the boundary as follows, 
\begin{align}
    \int_{x_{11},\tau}\delta g_{\mu\nu}\sim\frac{2\kappa^2_7 L^5}{6}\frac{\big<T_{\mu\nu}\big>}{r^4}+..., \label{boundaryEMT}
\end{align}
where $\kappa^2_7$ is related to the 11-dimensional gravitational constant as $\kappa^2_{11}={\rm Vol}(S^4)\kappa^2_7=8\pi^2/3(L/2)^4\kappa^2_7$.
Note that, to obtain the above, we need to apply the holographic renormalization in the M-theory, developed by Ref.~\cite{deHaro:2000vlm}. 
The reason for considering the 11-dimensional theory is that it is technically convenient to perform holographic renormalization. In the non-conformal 10-dimensional theory, holographic renormalization was developed in Ref.~\cite{Kanitscheider:2008kd}

The metric fluctuations $\delta g_{MN}$ of the bulk obey the following linearized Einstein equations:
\begin{align}
    \bar{\mathcal{H}}_{MN}=&\mathcal{H}_{MN}-\frac{g_{MN}}{2}\mathcal{H}^P_P=-2\kappa^2_7\mathcal{T}_{MN}, \label{linearEq} \\
    \mathcal{H}_{MN}\equiv& 
    \nabla^2\delta g_{MN}+\nabla_M\nabla_N\delta g^ P_P \notag \\
    &-\nabla^P(\nabla_M\delta g_{NP}+\nabla_N\delta g_{MP})-\frac{12}{L^2}\delta g_{MN}, \notag
\end{align}
where $M$, $N$ and $P$ are the Lorenz indices of the 7-dimensional space-time after integration over $S^4$, as in $ M,N,P=0,1,2,3,\tau,r,11$, and the source $\mathcal{T}_{MN}$ is the EMT of the matter fields which here considered only the pion field. 
We solve Eq.~\eqref{linearEq} under the axial gauge, $\delta g_{Mr} = 0$. 

In order to examine the EMT $\mathcal{T}_{MN}$ in the bulk, we embed the D8 branes into the 11-dimensional space, considering the action as
\begin{align}
    &S_{D8}= \notag \\
    &-\frac{C}{2\pi R_{11}g_{s}}\int d\tau dx_{11}d^4xdrd\Omega_4(\delta(\tau)+\delta(\tau-\pi/M_{KK})) \notag \\
    &\hspace{10mm}\times\frac{\sqrt{-G_{(11)}}}{\sqrt{G_{\tau\tau}}}\frac{1}{4}G^{MN}G^{PQ}{\rm tr}[F_{MP}F_{NQ}]+...,
\end{align}
with $C=(64\pi^6l_s^5)^{-1}$ and the metric tensor $G_{MN}$ in Eq.~\eqref{AdS7BH}. We define the EMT as
\begin{align}
    \mathcal{T}_{MN}=-\frac{2}{\sqrt{-G_{(11)}}}\frac{\delta S_{D8}}{\delta G^{MN}_{(11)}}{\rm Vol}(S^4) ,
\end{align}
which leads to the following expression:
\begin{widetext}
\begin{align}
    &\mathcal{T}_{MN}(\vec{\Delta},r) \equiv 2\pi R_{11}\int d\tau^\prime \mathcal{T}_{MN}(\tau^\prime,\vec{k},r), \notag \\
    &\mathcal{T}_{\mu\nu} (\vec{\Delta},r)=\frac{L^5}{{r}^5}\frac{\pi^2C{r}^2 L^2}{3g_s \sqrt{f(r))}}{\rm tr}\Big[F_{\mu\rho}{F_\nu}^{\rho}+(1+z^2)^{4/3}M_{KK}^2F_{\mu z}F_{\nu z}-\frac{\eta_{\mu\nu}}{4}\big(F_{\rho\sigma}F^{\rho\sigma}+2(1+z^2)^{4/3}M^2_{KK}F_{\rho z}{F^\rho}_z\big)\Big], \notag \\
    &\mathcal{T}_{11,11}(\vec{\Delta},r)=\frac{L^5}{{r}^5}\frac{\pi^2 Cr L^2}{3g_s\sqrt{f(r)}}{\rm tr}\Big[-\frac{1}{4}F_{\rho\sigma}F^{\rho\sigma}-\frac{(1+z^2)^{4/3}M^2_{KK}}{2}F_{\rho z}F^{\rho z}\Big], \notag \\
    &\mathcal{T}_{\mu z}(\vec{\Delta},r)=\frac{L^5}{{r}^5}\frac{\pi^2 Cr L^2}{3g_s\sqrt{f(r)}}{\rm tr}\Big[F_{\mu\rho}{F_z}^\rho\Big], \ \ \ \mathcal{T}_{zz}(\vec{\Delta},r)=\frac{L^5}{{r}^5}\frac{\pi^2 Cr L^2}{3g_s\sqrt{f(r)}}{\rm tr}\Big[\frac{1}{2}F_{z\rho}{F_z}^\rho-\frac{1}{4(1+z^2)^{4/3}M^2_{KK}}F_{\rho\sigma}F^{\rho\sigma}\Big], \label{bulkEMT}
\end{align}
\end{widetext}
where the radial coordinate $r \ (R\leq r<\infty)$ is related to $z \ (-\infty<z<\infty)$ as $z=\pm\sqrt{r^6/R^6-1}$.
In this study, we consider the case where the pion field is the only matter field as the source of the gravitational waves and decompose $A_{M}(x^\nu,z)$ as follows
\begin{align}
    &A_\mu(x^\nu,z)=0, \ \ \  A_z(x^\nu,z)=\pi(x^\mu)\phi_0(z), \label{pion}\\
    &\phi_0(z)=c_0(1+z^2)^{-1}, \notag
\end{align}
where $c_0^{-2}=2M^2_{KK}\kappa\int dz(1+z^2)\phi_0(z)^2$ is the normalization constant of the eigenfunction $\phi_0(z)$~\cite{Sakai:2004cn}. Since the pion field is an external line, it can be replaced by a plane wave as \(\pi(x^\nu)\sim e^{ip_\nu x^\nu}\) with incoming momentum $p_\nu$. 
More precisely, following the method discussed in Ref.~\cite{Abidin:2008hn}, we calculate the three-point function $\big<J_A^{\alpha}T^{\mu\nu}J_A^{\beta}\big>$ of the axial vector current $J_A^\alpha$ and the EMT $T^{\mu\nu}$ by the Gubser-Klebanov-Polyakov and Witten (GKP-Witten) relation \cite{Gubser:1998bc,Witten:1998qj} and extract only the pion states using the completeness relation of spectra of the axial vector meson.

Since we are interested in the EMT of QCD at the boundary such as to satisfy $\Delta^\mu\big<T_{\mu\nu}\big>=0$, we consider the metric fluctuations that are transverse $\Delta^\mu\delta g_{\mu\nu}\sim0$ near the boundary, which can be parametrized as $\delta g_{ij}\sim\delta g^{\rm TT}_{ij}+(\delta_{ij}-\Delta_i\Delta_j/\vec{\Delta}^2)/5\delta g^\alpha_\alpha$ with the transverse-traceless (TT) part $\delta g^{\rm TT}_{ij}$, $i,j=1,2,3$ and $\alpha=\mu,\tau,11$. Under the axial gauge, the Einstein equation requires the six-dimensional traceless condition $\delta g^\alpha_\alpha\sim0$ in the asymptotic AdS region ($r\rightarrow\infty$) labeled by $(x^\mu,\tau,x^{11})$. 
Therefore, we consider TT modes in the metric fluctuation, which is nothing but glueball spectra.
Among the 14 TT modes named in Ref.~\cite{Brower:2000rp} (see also Ref.~\cite{Constable:1999gb}), only $\rm T_{4}$ and $\rm S_{4}$ contribute to the EMT. Actually, substituting the pion field \eqref{pion} into Eq.~\eqref{gOther}, we confirm that $\delta g^{\rm other}_{\mu\nu}$ falls faster than $\rm T_4$ and $\rm S_4$ mode at $r\rightarrow\infty$. 
By decomposing the metric fluctuations into each mode, $\delta g_{\mu\nu}=\delta g^{\rm T(2)}_{\mu\nu}+\delta g^{\rm T(0)}_{\mu\nu}+\delta g^{\rm S(0)}_{\mu\nu}+\delta g^{\rm other}_{\mu\nu}$, the Einstein equations reduce to 
\begin{align}
    &\frac{r^2}{L^2}\Big[\frac{1}{L^2r^5}\partial_r\big((r^7-rR^6)\partial_r\big)-\frac{L^2\Delta^2}{r^2}\Big]\frac{L^2}{r^2}\delta g^{\rm T}_{\mu\nu}=-2\kappa^2_7\mathcal{T}^{\rm T}_{\mu\nu}, \\
    &\frac{r^2}{L^2}\Big[\frac{1}{L^2r^5}\partial_r\Big((r^7-rR^6)\Big(\partial_r+\frac{144r^5R^6}{(5r^5-2R^6)(r^6+2R^6)}\Big)\Big) \notag \\
    &\hspace{26mm}-\frac{L^2\Delta^2}{r^2}\Big]\frac{L^2}{r^2}\delta g^{\rm S(0)}_{\mu\nu}\Big|_{r\rightarrow\infty}=-2\kappa^2_7\mathcal{T}^{\rm S(0)}_{\mu\nu},
\end{align}
where the source of each mode is decomposed as $\mathcal{T}_{\mu\nu}=\mathcal{T}^{\rm T}_{\mu\nu}+\mathcal{T}^{\rm S(0)}_{\mu\nu}+\mathcal{T}^{\rm other}_{\mu\nu}$ $(\mathcal{T}^{\rm T}_{\mu\nu}\equiv\mathcal{T}^{\rm T(2)}_{\mu\nu}+\mathcal{T}^{\rm T(0)}_{\mu\nu})$, and the two $\rm T_4$-modes are combined as $\delta g^{\rm T}_{\mu\nu}=\delta g^{\rm T(2)}_{\mu\nu}+\delta g^{\rm T(0)}_{\mu\nu}$, since they obey the same equation. 

Following the same manner as in Ref.~\cite{Fujita:2022jus}, each part of EMT is written as 
\begin{widetext}
\begin{align}
    &2\kappa^2_7\mathcal{T}^{\rm T}_{\mu\nu}=2\kappa^2_7\mathcal{T}_{\mu\nu}-2\kappa^2_7\mathcal{T}^{\rm S}_{\mu\nu}-2\kappa^2_7\mathcal{T}^{\rm other}_{\mu\nu}, \\
    &2\kappa^2_7\mathcal{T}^{\rm S}_{\mu\nu}=\frac{r^6+2R^6}{4(r^6-R^6)}\Big(\eta_{\mu\nu}-\frac{\Delta_\mu \Delta_\nu}{\Delta^2}\Big)\Big[\frac{1}{L^4r^3}\partial_r\big(r(r^6-R^6)\partial_r(a+b)-3R^6(a+b)\big)-\Delta^2b\Big], \\ 
    &2\kappa^2_7\mathcal{T}^{\rm other}_{\mu\nu}=\frac{1}{L^4r^3}\partial_r\Big[\Big(\eta_{\mu\nu}-\frac{\Delta_\mu \Delta_\nu}{\Delta^2}\Big)r(r^6-R^6)\partial_r a-3R^6\eta_{\mu\nu}b\Big],  \\
    &\partial_r a=-\frac{3R^6}{(5r^6-2R^6)}b+\frac{9r^{11}}{(5r^6-2R^6)R^6}2\kappa^2_7\mathcal{T}_{zz}, \ \ \ b=-i\frac{\Delta^\mu}{\Delta^2}\frac{r^6\sqrt{r^6-R^6}}{R^9} 2\kappa^2_7 \mathcal{T}_{\mu z}, 
\end{align}
\end{widetext}
with the ansatz, 
\begin{align}
    &\delta g^{\rm other}_{\mu\nu}(\vec{\Delta},r)=\frac{r^2}{L^2}\frac{\Delta_\mu \Delta_\nu}{\Delta^2}a(r,\Delta), \notag \\
    &\delta g^{\rm other}_{11,11}(\vec{\Delta},r)=\frac{r^2}{L^2}b(r,\Delta), \notag \\
    &\delta g^{\rm other}_{\tau\tau}(\vec{\Delta},r)=\frac{r^2f(r)}{L^2}(-b(r,\Delta)). \label{gOther}
\end{align}

We determine the perturbation to the metric caused by the pion field as the bulk EMT source using Green's function method as follows, 
\begin{align}
    &\delta g^{\rm T/S}_{\mu\nu}(\vec{\Delta},r)\sim\frac{r^2}{L^2}\int dr^\prime \sqrt{-G_{(7)}(r^\prime)}G^{\rm T/S}(\vec{\Delta},r,r^\prime) \notag \\
    &\hspace{40mm}\times\frac{L^2}{{r^\prime}^2}2\kappa^2_7\mathcal{T}^{\rm T/S}_{\mu\nu}(\vec{\Delta},r^\prime), \label{solg}
\end{align}
where, for $\rm S_4$ mode, we can write as this form in the $r\rightarrow\infty$ limit, and $\sim$ implies this point. 
The Green's function $G^{\rm T/S}$ is written as 
\begin{align}
    G^{\rm T/S}(\vec{\Delta},r,r^\prime)=\sum^\infty_{n=1}\frac{\Psi^{\rm T/S}_n(r)\Psi^{\rm T/S}_n(r^\prime)}{(m^{\rm T/S}_n)^2+\vec{\Delta}^2},
\end{align}
by expanding in the complete system $\{\Psi^{\rm T/S}_n(r)\}$ which obeys the eigenvalue equation, 
\begin{align}
    &\frac{1}{r^3L^4}\partial_r\big((r^7-rR^6)\partial_r\Psi^{\rm T}_n\big)=-(m^{\rm T}_n)^2\Psi^{\rm T}_n, \\
    &\frac{1}{r^3L^4}\partial_r\Big((r^7-rR^6)\Big(\partial_r+\frac{144r^5R^6}{(5r^6-2R^6)(r^6+2R^6)}\Big)\Psi^{\rm S}_n\Big) \notag \\
    &\hspace{45mm}=-(m^{\rm S}_n)^2\Psi^{\rm S}_n, \label{EVeq}
\end{align}
where $m^{\rm T/S}_n$ is $n$th glueball mass of $\rm T/S$ mode and $\Psi^{\rm T/S}_n(r)$ satisfy the completeness conditions,
\begin{align}
    &\frac{r^3}{L^3}\sum^\infty_{n=1}\Psi^{\rm T/S}_n(r)\Psi^{\rm T/S}_n(r^\prime)=\delta(r-r^\prime), \notag \\
    &\int^\infty_R dr\frac{r^3}{L^3}\Psi^{\rm T/S}_n(r)\Psi^{\rm T/S}_m(r)=\delta_{mn}. \label{CR}
\end{align}

Note that the main components of the EMT in the bulk propagated by each of the T(2), T(0) and S(0) states are $\mathcal{T}_{\mu\nu}$, $\mathcal{T}_{11,11}$ and $\mathcal{T}_{\tau\tau}$ respectively. Even though the total $\mathcal{T}_{\tau\tau}$ is zero from Eq.~\eqref{bulkEMT}, the $\mathcal{T}_{\tau\tau}^{\rm S}$ satisfying $\mathcal{T}_{\tau\tau}^{\rm S}+\mathcal{T}_{\tau\tau}^{\rm other}=0$ is the source in the bulk propagated by the ${\rm S}_4$ mode.

\vspace{\baselineskip}

{\it Results.}---
Substituting the pion field into the EMT of the bulk, we read off the matrix elements of the boundary EMT from Eq.~\eqref{boundaryEMT} with the dictionary Eq.~\eqref{solg} as follows, 
\begin{align}
    &A(t)=\frac{6}{L^7}\int^\infty_Rdr^\prime\sum^\infty_{n=1}\frac{\alpha^{\rm T}_n\Psi^{\rm T}_n(r^\prime)}{(m^{\rm T}_n)^2+\vec{\Delta}^2}\frac{{r^\prime}^3}{L^3}P^{\mu\nu}_{\rm t}\mathcal{T}_{\mu\nu}(\vec{\Delta},r^\prime), \\
    &D(t)= \notag \\
    &-\frac{6}{L^7}\int^\infty_Rdr^\prime\sum^\infty_{n=1}\frac{\alpha^{\rm T}_n\Psi^{\rm T}_n(r^\prime)}{(m^{\rm T}_n)^2+\vec{\Delta}^2}\frac{{r^\prime}^3}{L^3}P^{\mu\nu}_{\rm s}(\mathcal{T}_{\mu\nu}+\mathcal{T}^{\rm S}_{\mu\nu}+\mathcal{T}^{\rm other}_{\mu\nu}) \notag \\
    &+\frac{6}{L^7}\int^\infty_Rdr^\prime\sum^\infty_{n=1}\frac{\alpha^{\rm S}_n\Psi^{\rm S}_n(r^\prime)}{(m^{\rm S}_n)^2+\vec{\Delta}^2}\frac{{r^\prime}^3}{L^3}P^{\mu\nu}_{\rm s}\mathcal{T}^{\rm S}_{\mu\nu}, \\
    &\alpha^{\rm T/S}=\frac{(m^{\rm T/S}_n)^2L^4}{6}\int^\infty_R dr r^3\Psi^{\rm T/S}_n(r),
\end{align} 
where $P^{\mu\nu}_{\rm t}$ and $P^{\mu\nu}_{\rm s}$ are defined as $P^{\mu\nu}_{\rm t}=(3P^\mu P^\nu/P^2-\eta^{\mu\nu})/P^2$ and $P^{\mu\nu}_{\rm s}=(P^\mu P^\nu/P^2-\eta^{\mu\nu})/\Delta^2$. 

To calculate infinite sums, we expand as 
\begin{align}
    &\frac{6}{L^7}\int^\infty_Rdr^\prime\sum^\infty_{n=1}\frac{\alpha^{\rm T/S}_n\Psi^{\rm T/S}_n(r^\prime)}{(m^{\rm T/S}_n)^2+\vec{\Delta}^2}=\sum^\infty_{k=0}F_k^{\rm T/S}(r^\prime)(-\vec{\Delta}^2)^k, \notag \\
    &F^{\rm T/S}_k(r^\prime)=\int_{R}^{\infty} dr\frac{r^3}{L^3}\sum_{n=0}^{\infty} \frac{\Psi_{n}^{\rm T/S}(r)\Psi_{n}^{\rm T/S}(r^{\prime})}{(m_{n})^{2k}}. \label{expand}
\end{align}
From the eigenvalue equations \eqref{EVeq}, $F^{\rm T/S}_k(r)$ satisfies the following equations:
\begin{align}
    &\frac{1}{r^3L^4}\partial_r\big((r^7-rR^6)\partial_rF^{\rm T}_k\big)=-F^{\rm T}_{k-1}, \\
    &\frac{1}{r^3L^4}\partial_r\Big((r^7-rR^6)\Big(\partial_r+\frac{144r^5R^6}{(5r^6-2R^6)(r^6+2R^6)}\Big)F^{\rm S}_k\Big) \notag \\
    &\hspace{60mm}=-F^{\rm S}_{k-1},
\end{align}
where we can prove $F^{\rm T/S}_0=1$ by using Eq.~\eqref{CR}. 
We determine $F^{\rm T/S}_k(r)$ by solving it as the recurrence formula with boundary conditions on $\partial_r F^{\rm T/S}_k(r)|_{r\rightarrow R}=1, \ F^{\rm T/S}_k(r)|_{r\rightarrow \infty}=0$. 

\begin{figure}
    \includegraphics[scale=0.25]{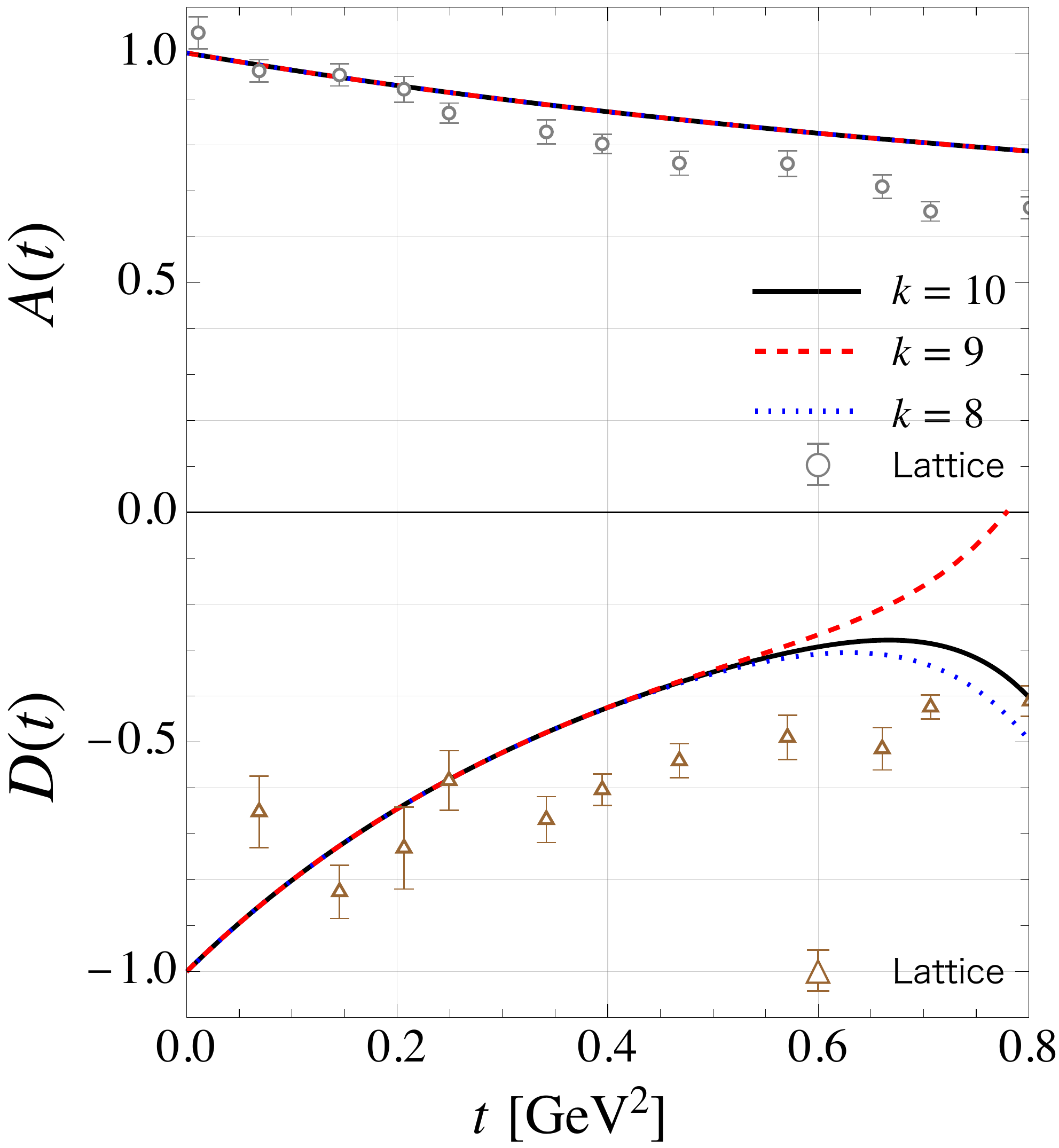}
    \caption{The pion GFFs $A(t)$ and $D(t)$ with the lattice data~\cite{Hackett:2023nkr}.}
    \label{AtDt}
\end{figure}

We show the results of GFFs that truncate the expansion at $k=10$ using Eq.~\eqref{expand} in FIG.~\ref{AtDt} (also including results that truncate at $k=8,9$), where we set $\kappa =0.00745$ and $M_{KK} =949 \, \rm{MeV}$ which are determined by $\rho$ meson mass $776 \ {\rm MeV}$ and the pion decay constant $f_{\pi}=92.4 \, \rm{MeV}$~\cite{Sakai:2005yt}. 
In the current calculation, $D(t)$ begins to deviate from around $t\sim0.5 \ {\rm GeV^2}$ while $A(t)$ converges well in the range $0<t<0.8 \ {\rm GeV^2}$ on the expansion up to $k=10$ with momentum transfer. The current approach begins to deviate from QCD in the region where $t$ is close to the $M_{KK}$, so further efforts are unnecessary. 

In the top-down holographic QCD approach, the amplitude of $D(t)$ dumps more rapidly than that of $A(t)$, which is qualitatively consistent with the results of lattice QCD~\cite{Hackett:2023nkr}. 
These differences are due to the contribution of $S_{4}$ modes to $D(t)$, which is not observed in the bottom-up holographic QCD framework~\cite{Abidin:2008hn}. 
This is because the usual bottom-up model, as used in Ref.~\cite{Abidin:2008hn}, does not have ${\rm S}_4$ mode with $\mathcal{T}_{\tau\tau}$ as source. 
Also in this approach, as in previous studies~\cite{Fujita:2022jus}, we observe the so-called glueball dominance, in which pions have gravitational interactions via infinite glueball spectra. 

As you can see in FIG.~\ref{AtDt}, our result shows $D(0)=-1$. 
Note that this is a result of our calculations with loop effects in the large $Nc$ limit, unlike the fact $ A(0)=1$, given by the general constraints. 
This is consistent with the low-energy theorem obtained in the soft pion limit~\cite{Donoghue:1991qv,Polyakov:1999gs,Hudson:2017xug}. The result is as expected since in our present work we consider the Sakai-Sugimoto model, which is the holographic dual of QCD in the chiral limit. Several methods have been proposed on how to introduce quark masses into this model~\cite{Casero:2007ae,Hashimoto:2007fa,Bergman:2007pm,Dhar:2007bz,Hashimoto:2008sr,Dhar:2008um,McNees:2008km}, thus analysis beyond the chiral limit is a future work.

\vspace{\baselineskip}

{\it Summary and outlook.}---
We calculated the GFFs of pions from the top-down holographic QCD. 
We solve the Einstein equations in bulk theory with pion fields as matter fields and derived the GFFs using the dictionary~\eqref{boundaryEMT}.
Our results are qualitatively compatible with those from lattice QCD~\cite{Hackett:2023nkr} and successfully reproduce the feature where the behaviors of $A(t)$ and $D(t)$ are different, which is distinct from bottom-up holographic QCD~\cite{Abidin:2008hn}.
Additionally, we obtained the D-term as $D(0)=-1$ in the chiral limit. 

Finally, this study opens up several avenues for future research. First, it is possible to generalize to other hadrons, such as nucleons and vector mesons with non-zero spin. Also, it is very interesting to investigate beyond the chiral limit, because the connection between the Gell-Mann--Oakes--Renner relation and the stability of the stress distribution has been discussed in Ref.~\cite{Son:2014sna}. It is also important to investigate the behavior of the GFFs and the stress distribution inside hadrons under finite temperatures, densities, and magnetic fields, which change the nature of the interaction in the system. 

\vspace{\baselineskip}

{\it Acknowledgements.}---
We are grateful to the authors of Ref.~\cite{Hackett:2023nkr} for generously supplying the data tables utilized in FIG.~\ref{AtDt}.
We also thank Masayasu Harada, Atsushi Hosaka, Shin Nakamura, Shigeki Sugimoto, Kei Suzuki, and Yasuhiro Yamaguchi for useful discussions. This work was supported by the Japan Society for the Promotion of Science (JSPS) KAKENHI (Grants No. JP24K17054, No. JP24KJ1620).

\nocite{*}


\bibliography{reference}

\begin{thebibliography}{116}%
\makeatletter
\providecommand \@ifxundefined [1]{%
 \@ifx{#1\undefined}
}%
\providecommand \@ifnum [1]{%
 \ifnum #1\expandafter \@firstoftwo
 \else \expandafter \@secondoftwo
 \fi
}%
\providecommand \@ifx [1]{%
 \ifx #1\expandafter \@firstoftwo
 \else \expandafter \@secondoftwo
 \fi
}%
\providecommand \natexlab [1]{#1}%
\providecommand \enquote  [1]{``#1''}%
\providecommand \bibnamefont  [1]{#1}%
\providecommand \bibfnamefont [1]{#1}%
\providecommand \citenamefont [1]{#1}%
\providecommand \href@noop [0]{\@secondoftwo}%
\providecommand \href [0]{\begingroup \@sanitize@url \@href}%
\providecommand \@href[1]{\@@startlink{#1}\@@href}%
\providecommand \@@href[1]{\endgroup#1\@@endlink}%
\providecommand \@sanitize@url [0]{\catcode `\\12\catcode `\$12\catcode
  `\&12\catcode `\#12\catcode `\^12\catcode `\_12\catcode `\%12\relax}%
\providecommand \@@startlink[1]{}%
\providecommand \@@endlink[0]{}%
\providecommand \url  [0]{\begingroup\@sanitize@url \@url }%
\providecommand \@url [1]{\endgroup\@href {#1}{\urlprefix }}%
\providecommand \urlprefix  [0]{URL }%
\providecommand \Eprint [0]{\href }%
\providecommand \doibase [0]{http://dx.doi.org/}%
\providecommand \selectlanguage [0]{\@gobble}%
\providecommand \bibinfo  [0]{\@secondoftwo}%
\providecommand \bibfield  [0]{\@secondoftwo}%
\providecommand \translation [1]{[#1]}%
\providecommand \BibitemOpen [0]{}%
\providecommand \bibitemStop [0]{}%
\providecommand \bibitemNoStop [0]{.\EOS\space}%
\providecommand \EOS [0]{\spacefactor3000\relax}%
\providecommand \BibitemShut  [1]{\csname bibitem#1\endcsname}%
\let\auto@bib@innerbib\@empty
\bibitem [{\citenamefont {Polyakov}\ and\ \citenamefont
  {Schweitzer}(2018)}]{Polyakov:2018zvc}%
  \BibitemOpen
  \bibfield  {author} {\bibinfo {author} {\bibfnamefont {Maxim~V.}\
  \bibnamefont {Polyakov}}\ and\ \bibinfo {author} {\bibfnamefont {Peter}\
  \bibnamefont {Schweitzer}},\ }\bibfield  {title} {\enquote {\bibinfo {title}
  {{Forces inside hadrons: pressure, surface tension, mechanical radius, and
  all that}},}\ }\href {\doibase 10.1142/S0217751X18300259} {\bibfield
  {journal} {\bibinfo  {journal} {Int. J. Mod. Phys. A}\ }\textbf {\bibinfo
  {volume} {33}},\ \bibinfo {pages} {1830025} (\bibinfo {year} {2018})},\
  \Eprint {http://arxiv.org/abs/1805.06596} {arXiv:1805.06596 [hep-ph]}
  \BibitemShut {NoStop}%
\bibitem [{\citenamefont {Burkert}\ \emph {et~al.}(2023)\citenamefont
  {Burkert}, \citenamefont {Elouadrhiri}, \citenamefont {Girod}, \citenamefont
  {Lorc\'e}, \citenamefont {Schweitzer},\ and\ \citenamefont
  {Shanahan}}]{Burkert:2023wzr}%
  \BibitemOpen
  \bibfield  {author} {\bibinfo {author} {\bibfnamefont {V.~D.}\ \bibnamefont
  {Burkert}}, \bibinfo {author} {\bibfnamefont {L.}~\bibnamefont
  {Elouadrhiri}}, \bibinfo {author} {\bibfnamefont {F.~X.}\ \bibnamefont
  {Girod}}, \bibinfo {author} {\bibfnamefont {C.}~\bibnamefont {Lorc\'e}},
  \bibinfo {author} {\bibfnamefont {P.}~\bibnamefont {Schweitzer}}, \ and\
  \bibinfo {author} {\bibfnamefont {P.~E.}\ \bibnamefont {Shanahan}},\
  }\bibfield  {title} {\enquote {\bibinfo {title} {{Colloquium: Gravitational
  form factors of the proton}},}\ }\href {\doibase
  10.1103/RevModPhys.95.041002} {\bibfield  {journal} {\bibinfo  {journal}
  {Rev. Mod. Phys.}\ }\textbf {\bibinfo {volume} {95}},\ \bibinfo {pages}
  {041002} (\bibinfo {year} {2023})},\ \Eprint
  {http://arxiv.org/abs/2303.08347} {arXiv:2303.08347 [hep-ph]} \BibitemShut
  {NoStop}%
\bibitem [{\citenamefont {Diehl}(2003)}]{Diehl:2003ny}%
  \BibitemOpen
  \bibfield  {author} {\bibinfo {author} {\bibfnamefont {M.}~\bibnamefont
  {Diehl}},\ }\bibfield  {title} {\enquote {\bibinfo {title} {{Generalized
  parton distributions}},}\ }\href {\doibase 10.1016/j.physrep.2003.08.002}
  {\bibfield  {journal} {\bibinfo  {journal} {Phys. Rept.}\ }\textbf {\bibinfo
  {volume} {388}},\ \bibinfo {pages} {41--277} (\bibinfo {year} {2003})},\
  \Eprint {http://arxiv.org/abs/hep-ph/0307382} {arXiv:hep-ph/0307382}
  \BibitemShut {NoStop}%
\bibitem [{\citenamefont {Belitsky}\ and\ \citenamefont
  {Radyushkin}(2005)}]{Belitsky:2005qn}%
  \BibitemOpen
  \bibfield  {author} {\bibinfo {author} {\bibfnamefont {A.~V.}\ \bibnamefont
  {Belitsky}}\ and\ \bibinfo {author} {\bibfnamefont {A.~V.}\ \bibnamefont
  {Radyushkin}},\ }\bibfield  {title} {\enquote {\bibinfo {title} {{Unraveling
  hadron structure with generalized parton distributions}},}\ }\href {\doibase
  10.1016/j.physrep.2005.06.002} {\bibfield  {journal} {\bibinfo  {journal}
  {Phys. Rept.}\ }\textbf {\bibinfo {volume} {418}},\ \bibinfo {pages} {1--387}
  (\bibinfo {year} {2005})},\ \Eprint {http://arxiv.org/abs/hep-ph/0504030}
  {arXiv:hep-ph/0504030} \BibitemShut {NoStop}%
\bibitem [{\citenamefont {Mezrag}(2022)}]{Mezrag:2022pqk}%
  \BibitemOpen
  \bibfield  {author} {\bibinfo {author} {\bibfnamefont {C\'edric}\
  \bibnamefont {Mezrag}},\ }\bibfield  {title} {\enquote {\bibinfo {title} {{An
  Introductory Lecture on Generalised Parton Distributions}},}\ }\href
  {\doibase 10.1007/s00601-022-01765-x} {\bibfield  {journal} {\bibinfo
  {journal} {Few Body Syst.}\ }\textbf {\bibinfo {volume} {63}},\ \bibinfo
  {pages} {62} (\bibinfo {year} {2022})},\ \Eprint
  {http://arxiv.org/abs/2207.13584} {arXiv:2207.13584 [hep-ph]} \BibitemShut
  {NoStop}%
\bibitem [{\citenamefont {Burkert}\ \emph {et~al.}(2018)\citenamefont
  {Burkert}, \citenamefont {Elouadrhiri},\ and\ \citenamefont
  {Girod}}]{Burkert:2018bqq}%
  \BibitemOpen
  \bibfield  {author} {\bibinfo {author} {\bibfnamefont {V.~D.}\ \bibnamefont
  {Burkert}}, \bibinfo {author} {\bibfnamefont {L.}~\bibnamefont
  {Elouadrhiri}}, \ and\ \bibinfo {author} {\bibfnamefont {F.~X.}\ \bibnamefont
  {Girod}},\ }\bibfield  {title} {\enquote {\bibinfo {title} {{The pressure
  distribution inside the proton}},}\ }\href {\doibase
  10.1038/s41586-018-0060-z} {\bibfield  {journal} {\bibinfo  {journal}
  {Nature}\ }\textbf {\bibinfo {volume} {557}},\ \bibinfo {pages} {396--399}
  (\bibinfo {year} {2018})}\BibitemShut {NoStop}%
\bibitem [{\citenamefont {Duran}\ \emph {et~al.}(2023)\citenamefont {Duran}
  \emph {et~al.}}]{Duran:2022xag}%
  \BibitemOpen
  \bibfield  {author} {\bibinfo {author} {\bibfnamefont {B.}~\bibnamefont
  {Duran}} \emph {et~al.},\ }\bibfield  {title} {\enquote {\bibinfo {title}
  {{Determining the gluonic gravitational form factors of the proton}},}\
  }\href {\doibase 10.1038/s41586-023-05730-4} {\bibfield  {journal} {\bibinfo
  {journal} {Nature}\ }\textbf {\bibinfo {volume} {615}},\ \bibinfo {pages}
  {813--816} (\bibinfo {year} {2023})},\ \Eprint
  {http://arxiv.org/abs/2207.05212} {arXiv:2207.05212 [nucl-ex]} \BibitemShut
  {NoStop}%
\bibitem [{\citenamefont {Uehara}\ \emph {et~al.}(2012)\citenamefont {Uehara}
  \emph {et~al.}}]{Belle:2012wwz}%
  \BibitemOpen
  \bibfield  {author} {\bibinfo {author} {\bibfnamefont {S.}~\bibnamefont
  {Uehara}} \emph {et~al.} (\bibinfo {collaboration} {Belle}),\ }\bibfield
  {title} {\enquote {\bibinfo {title} {{Measurement of $\gamma \gamma^* \to
  \pi^0$ transition form factor at Belle}},}\ }\href {\doibase
  10.1103/PhysRevD.86.092007} {\bibfield  {journal} {\bibinfo  {journal} {Phys.
  Rev. D}\ }\textbf {\bibinfo {volume} {86}},\ \bibinfo {pages} {092007}
  (\bibinfo {year} {2012})},\ \Eprint {http://arxiv.org/abs/1205.3249}
  {arXiv:1205.3249 [hep-ex]} \BibitemShut {NoStop}%
\bibitem [{\citenamefont {Masuda}\ \emph {et~al.}(2016)\citenamefont {Masuda}
  \emph {et~al.}}]{Belle:2015oin}%
  \BibitemOpen
  \bibfield  {author} {\bibinfo {author} {\bibfnamefont {M.}~\bibnamefont
  {Masuda}} \emph {et~al.} (\bibinfo {collaboration} {Belle}),\ }\bibfield
  {title} {\enquote {\bibinfo {title} {{Study of $\pi^0$ pair production in
  single-tag two-photon collisions}},}\ }\href {\doibase
  10.1103/PhysRevD.93.032003} {\bibfield  {journal} {\bibinfo  {journal} {Phys.
  Rev. D}\ }\textbf {\bibinfo {volume} {93}},\ \bibinfo {pages} {032003}
  (\bibinfo {year} {2016})},\ \Eprint {http://arxiv.org/abs/1508.06757}
  {arXiv:1508.06757 [hep-ex]} \BibitemShut {NoStop}%
\bibitem [{\citenamefont {Kumano}\ \emph {et~al.}(2018)\citenamefont {Kumano},
  \citenamefont {Song},\ and\ \citenamefont {Teryaev}}]{Kumano:2017lhr}%
  \BibitemOpen
  \bibfield  {author} {\bibinfo {author} {\bibfnamefont {S.}~\bibnamefont
  {Kumano}}, \bibinfo {author} {\bibfnamefont {Qin-Tao}\ \bibnamefont {Song}},
  \ and\ \bibinfo {author} {\bibfnamefont {O.~V.}\ \bibnamefont {Teryaev}},\
  }\bibfield  {title} {\enquote {\bibinfo {title} {{Hadron tomography by
  generalized distribution amplitudes in pion-pair production process $\gamma^*
  \gamma \rightarrow \pi^0 \pi^0 $ and gravitational form factors for pion}},}\
  }\href {\doibase 10.1103/PhysRevD.97.014020} {\bibfield  {journal} {\bibinfo
  {journal} {Phys. Rev. D}\ }\textbf {\bibinfo {volume} {97}},\ \bibinfo
  {pages} {014020} (\bibinfo {year} {2018})},\ \Eprint
  {http://arxiv.org/abs/1711.08088} {arXiv:1711.08088 [hep-ph]} \BibitemShut
  {NoStop}%
\bibitem [{\citenamefont {Hagler}\ \emph {et~al.}(2008)\citenamefont {Hagler}
  \emph {et~al.}}]{LHPC:2007blg}%
  \BibitemOpen
  \bibfield  {author} {\bibinfo {author} {\bibfnamefont {Ph.}\ \bibnamefont
  {Hagler}} \emph {et~al.} (\bibinfo {collaboration} {LHPC}),\ }\bibfield
  {title} {\enquote {\bibinfo {title} {{Nucleon Generalized Parton
  Distributions from Full Lattice QCD}},}\ }\href {\doibase
  10.1103/PhysRevD.77.094502} {\bibfield  {journal} {\bibinfo  {journal} {Phys.
  Rev. D}\ }\textbf {\bibinfo {volume} {77}},\ \bibinfo {pages} {094502}
  (\bibinfo {year} {2008})},\ \Eprint {http://arxiv.org/abs/0705.4295}
  {arXiv:0705.4295 [hep-lat]} \BibitemShut {NoStop}%
\bibitem [{\citenamefont {Bali}\ \emph {et~al.}(2016)\citenamefont {Bali},
  \citenamefont {Collins}, \citenamefont {G\"ockeler}, \citenamefont {R\"odl},
  \citenamefont {Sch\"afer},\ and\ \citenamefont {Sternbeck}}]{Bali:2016wqg}%
  \BibitemOpen
  \bibfield  {author} {\bibinfo {author} {\bibfnamefont {Gunnar}\ \bibnamefont
  {Bali}}, \bibinfo {author} {\bibfnamefont {Sara}\ \bibnamefont {Collins}},
  \bibinfo {author} {\bibfnamefont {Meinulf}\ \bibnamefont {G\"ockeler}},
  \bibinfo {author} {\bibfnamefont {Rudolf}\ \bibnamefont {R\"odl}}, \bibinfo
  {author} {\bibfnamefont {Andreas}\ \bibnamefont {Sch\"afer}}, \ and\ \bibinfo
  {author} {\bibfnamefont {Andre}\ \bibnamefont {Sternbeck}},\ }\bibfield
  {title} {\enquote {\bibinfo {title} {{Nucleon generalized form factors from
  lattice QCD with nearly physical quark masses}},}\ }\href {\doibase
  10.22323/1.251.0118} {\bibfield  {journal} {\bibinfo  {journal} {PoS}\
  }\textbf {\bibinfo {volume} {LATTICE2015}},\ \bibinfo {pages} {118} (\bibinfo
  {year} {2016})},\ \Eprint {http://arxiv.org/abs/1601.04818} {arXiv:1601.04818
  [hep-lat]} \BibitemShut {NoStop}%
\bibitem [{\citenamefont {Shanahan}\ and\ \citenamefont
  {Detmold}(2019)}]{Shanahan:2018pib}%
  \BibitemOpen
  \bibfield  {author} {\bibinfo {author} {\bibfnamefont {P.~E.}\ \bibnamefont
  {Shanahan}}\ and\ \bibinfo {author} {\bibfnamefont {W.}~\bibnamefont
  {Detmold}},\ }\bibfield  {title} {\enquote {\bibinfo {title} {{Gluon
  gravitational form factors of the nucleon and the pion from lattice QCD}},}\
  }\href {\doibase 10.1103/PhysRevD.99.014511} {\bibfield  {journal} {\bibinfo
  {journal} {Phys. Rev. D}\ }\textbf {\bibinfo {volume} {99}},\ \bibinfo
  {pages} {014511} (\bibinfo {year} {2019})},\ \Eprint
  {http://arxiv.org/abs/1810.04626} {arXiv:1810.04626 [hep-lat]} \BibitemShut
  {NoStop}%
\bibitem [{\citenamefont {Alexandrou}\ \emph {et~al.}(2018)\citenamefont
  {Alexandrou}, \citenamefont {Constantinou}, \citenamefont {Hadjiyiannakou},
  \citenamefont {Jansen}, \citenamefont {Kallidonis}, \citenamefont {Koutsou},\
  and\ \citenamefont {Vaquero Avil\'es-Casco}}]{Alexandrou:2018xnp}%
  \BibitemOpen
  \bibfield  {author} {\bibinfo {author} {\bibfnamefont {Constantia}\
  \bibnamefont {Alexandrou}}, \bibinfo {author} {\bibfnamefont {Martha}\
  \bibnamefont {Constantinou}}, \bibinfo {author} {\bibfnamefont {Kyriakos}\
  \bibnamefont {Hadjiyiannakou}}, \bibinfo {author} {\bibfnamefont {Karl}\
  \bibnamefont {Jansen}}, \bibinfo {author} {\bibfnamefont {Christos}\
  \bibnamefont {Kallidonis}}, \bibinfo {author} {\bibfnamefont {Giannis}\
  \bibnamefont {Koutsou}}, \ and\ \bibinfo {author} {\bibfnamefont {Alejandro}\
  \bibnamefont {Vaquero Avil\'es-Casco}},\ }\bibfield  {title} {\enquote
  {\bibinfo {title} {{Nucleon spin structure from lattice QCD}},}\ }\href
  {\doibase 10.22323/1.316.0148} {\bibfield  {journal} {\bibinfo  {journal}
  {PoS}\ }\textbf {\bibinfo {volume} {DIS2018}},\ \bibinfo {pages} {148}
  (\bibinfo {year} {2018})},\ \Eprint {http://arxiv.org/abs/1807.11214}
  {arXiv:1807.11214 [hep-lat]} \BibitemShut {NoStop}%
\bibitem [{\citenamefont {Alexandrou}\ \emph
  {et~al.}(2020{\natexlab{a}})\citenamefont {Alexandrou} \emph
  {et~al.}}]{Alexandrou:2019ali}%
  \BibitemOpen
  \bibfield  {author} {\bibinfo {author} {\bibfnamefont {C.}~\bibnamefont
  {Alexandrou}} \emph {et~al.},\ }\bibfield  {title} {\enquote {\bibinfo
  {title} {{Moments of nucleon generalized parton distributions from lattice
  QCD simulations at physical pion mass}},}\ }\href {\doibase
  10.1103/PhysRevD.101.034519} {\bibfield  {journal} {\bibinfo  {journal}
  {Phys. Rev. D}\ }\textbf {\bibinfo {volume} {101}},\ \bibinfo {pages}
  {034519} (\bibinfo {year} {2020}{\natexlab{a}})},\ \Eprint
  {http://arxiv.org/abs/1908.10706} {arXiv:1908.10706 [hep-lat]} \BibitemShut
  {NoStop}%
\bibitem [{\citenamefont {Alexandrou}\ \emph
  {et~al.}(2020{\natexlab{b}})\citenamefont {Alexandrou}, \citenamefont
  {Bacchio}, \citenamefont {Constantinou}, \citenamefont {Finkenrath},
  \citenamefont {Hadjiyiannakou}, \citenamefont {Jansen}, \citenamefont
  {Koutsou}, \citenamefont {Panagopoulos},\ and\ \citenamefont
  {Spanoudes}}]{Alexandrou:2020sml}%
  \BibitemOpen
  \bibfield  {author} {\bibinfo {author} {\bibfnamefont {C.}~\bibnamefont
  {Alexandrou}}, \bibinfo {author} {\bibfnamefont {S.}~\bibnamefont {Bacchio}},
  \bibinfo {author} {\bibfnamefont {M.}~\bibnamefont {Constantinou}}, \bibinfo
  {author} {\bibfnamefont {J.}~\bibnamefont {Finkenrath}}, \bibinfo {author}
  {\bibfnamefont {K.}~\bibnamefont {Hadjiyiannakou}}, \bibinfo {author}
  {\bibfnamefont {K.}~\bibnamefont {Jansen}}, \bibinfo {author} {\bibfnamefont
  {G.}~\bibnamefont {Koutsou}}, \bibinfo {author} {\bibfnamefont
  {H.}~\bibnamefont {Panagopoulos}}, \ and\ \bibinfo {author} {\bibfnamefont
  {G.}~\bibnamefont {Spanoudes}},\ }\bibfield  {title} {\enquote {\bibinfo
  {title} {{Complete flavor decomposition of the spin and momentum fraction of
  the proton using lattice QCD simulations at physical pion mass}},}\ }\href
  {\doibase 10.1103/PhysRevD.101.094513} {\bibfield  {journal} {\bibinfo
  {journal} {Phys. Rev. D}\ }\textbf {\bibinfo {volume} {101}},\ \bibinfo
  {pages} {094513} (\bibinfo {year} {2020}{\natexlab{b}})},\ \Eprint
  {http://arxiv.org/abs/2003.08486} {arXiv:2003.08486 [hep-lat]} \BibitemShut
  {NoStop}%
\bibitem [{\citenamefont {Pefkou}\ \emph {et~al.}(2022)\citenamefont {Pefkou},
  \citenamefont {Hackett},\ and\ \citenamefont {Shanahan}}]{Pefkou:2021fni}%
  \BibitemOpen
  \bibfield  {author} {\bibinfo {author} {\bibfnamefont {Dimitra~A.}\
  \bibnamefont {Pefkou}}, \bibinfo {author} {\bibfnamefont {Daniel~C.}\
  \bibnamefont {Hackett}}, \ and\ \bibinfo {author} {\bibfnamefont {Phiala~E.}\
  \bibnamefont {Shanahan}},\ }\bibfield  {title} {\enquote {\bibinfo {title}
  {{Gluon gravitational structure of hadrons of different spin}},}\ }\href
  {\doibase 10.1103/PhysRevD.105.054509} {\bibfield  {journal} {\bibinfo
  {journal} {Phys. Rev. D}\ }\textbf {\bibinfo {volume} {105}},\ \bibinfo
  {pages} {054509} (\bibinfo {year} {2022})},\ \Eprint
  {http://arxiv.org/abs/2107.10368} {arXiv:2107.10368 [hep-lat]} \BibitemShut
  {NoStop}%
\bibitem [{\citenamefont {Hackett}\ \emph {et~al.}(2024)\citenamefont
  {Hackett}, \citenamefont {Pefkou},\ and\ \citenamefont
  {Shanahan}}]{Hackett:2023rif}%
  \BibitemOpen
  \bibfield  {author} {\bibinfo {author} {\bibfnamefont {Daniel~C.}\
  \bibnamefont {Hackett}}, \bibinfo {author} {\bibfnamefont {Dimitra~A.}\
  \bibnamefont {Pefkou}}, \ and\ \bibinfo {author} {\bibfnamefont {Phiala~E.}\
  \bibnamefont {Shanahan}},\ }\bibfield  {title} {\enquote {\bibinfo {title}
  {{Gravitational Form Factors of the Proton from Lattice QCD}},}\ }\href
  {\doibase 10.1103/PhysRevLett.132.251904} {\bibfield  {journal} {\bibinfo
  {journal} {Phys. Rev. Lett.}\ }\textbf {\bibinfo {volume} {132}},\ \bibinfo
  {pages} {251904} (\bibinfo {year} {2024})},\ \Eprint
  {http://arxiv.org/abs/2310.08484} {arXiv:2310.08484 [hep-lat]} \BibitemShut
  {NoStop}%
\bibitem [{\citenamefont {Dutrieux}\ \emph {et~al.}(2024)\citenamefont
  {Dutrieux}, \citenamefont {Edwards}, \citenamefont {Egerer}, \citenamefont
  {Karpie}, \citenamefont {Monahan}, \citenamefont {Orginos}, \citenamefont
  {Radyushkin}, \citenamefont {Richards}, \citenamefont {Romero},\ and\
  \citenamefont {Zafeiropoulos}}]{HadStruc:2024rix}%
  \BibitemOpen
  \bibfield  {author} {\bibinfo {author} {\bibfnamefont {Herv\'e}\ \bibnamefont
  {Dutrieux}}, \bibinfo {author} {\bibfnamefont {Robert~G.}\ \bibnamefont
  {Edwards}}, \bibinfo {author} {\bibfnamefont {Colin}\ \bibnamefont {Egerer}},
  \bibinfo {author} {\bibfnamefont {Joseph}\ \bibnamefont {Karpie}}, \bibinfo
  {author} {\bibfnamefont {Christopher}\ \bibnamefont {Monahan}}, \bibinfo
  {author} {\bibfnamefont {Kostas}\ \bibnamefont {Orginos}}, \bibinfo {author}
  {\bibfnamefont {Anatoly}\ \bibnamefont {Radyushkin}}, \bibinfo {author}
  {\bibfnamefont {David}\ \bibnamefont {Richards}}, \bibinfo {author}
  {\bibfnamefont {Eloy}\ \bibnamefont {Romero}}, \ and\ \bibinfo {author}
  {\bibfnamefont {Savvas}\ \bibnamefont {Zafeiropoulos}} (\bibinfo
  {collaboration} {HadStruc}),\ }\bibfield  {title} {\enquote {\bibinfo {title}
  {{Towards Unpolarized GPDs from Pseudo-Distributions}},}\ }\href@noop {} {\
  (\bibinfo {year} {2024})},\ \Eprint {http://arxiv.org/abs/2405.10304}
  {arXiv:2405.10304 [hep-lat]} \BibitemShut {NoStop}%
\bibitem [{\citenamefont {Brommel}\ \emph {et~al.}(2006)\citenamefont
  {Brommel}, \citenamefont {Diehl}, \citenamefont {Gockeler}, \citenamefont
  {Hagler}, \citenamefont {Horsley}, \citenamefont {Pleiter}, \citenamefont
  {Rakow}, \citenamefont {Schafer}, \citenamefont {Schierholz},\ and\
  \citenamefont {Zanotti}}]{Brommel:2005ee}%
  \BibitemOpen
  \bibfield  {author} {\bibinfo {author} {\bibfnamefont {D.}~\bibnamefont
  {Brommel}}, \bibinfo {author} {\bibfnamefont {M.}~\bibnamefont {Diehl}},
  \bibinfo {author} {\bibfnamefont {M.}~\bibnamefont {Gockeler}}, \bibinfo
  {author} {\bibfnamefont {Ph.}\ \bibnamefont {Hagler}}, \bibinfo {author}
  {\bibfnamefont {R.}~\bibnamefont {Horsley}}, \bibinfo {author} {\bibfnamefont
  {D.}~\bibnamefont {Pleiter}}, \bibinfo {author} {\bibfnamefont {Paul E.~L.}\
  \bibnamefont {Rakow}}, \bibinfo {author} {\bibfnamefont {A.}~\bibnamefont
  {Schafer}}, \bibinfo {author} {\bibfnamefont {G.}~\bibnamefont {Schierholz}},
  \ and\ \bibinfo {author} {\bibfnamefont {J.~M.}\ \bibnamefont {Zanotti}},\
  }\bibfield  {title} {\enquote {\bibinfo {title} {{Structure of the pion from
  full lattice QCD}},}\ }\href {\doibase 10.22323/1.020.0360} {\bibfield
  {journal} {\bibinfo  {journal} {PoS}\ }\textbf {\bibinfo {volume}
  {LAT2005}},\ \bibinfo {pages} {360} (\bibinfo {year} {2006})},\ \Eprint
  {http://arxiv.org/abs/hep-lat/0509133} {arXiv:hep-lat/0509133} \BibitemShut
  {NoStop}%
\bibitem [{\citenamefont {Brommel}(2007)}]{Brommel:2007zz}%
  \BibitemOpen
  \bibfield  {author} {\bibinfo {author} {\bibfnamefont {Dirk}\ \bibnamefont
  {Brommel}},\ }\emph {\bibinfo {title} {{Pion Structure from the Lattice}}},\
  \href {\doibase 10.3204/DESY-THESIS-2007-023} {Ph.D. thesis},\ \bibinfo
  {school} {Regensburg U.} (\bibinfo {year} {2007})\BibitemShut {NoStop}%
\bibitem [{\citenamefont {Yang}\ \emph {et~al.}(2015)\citenamefont {Yang},
  \citenamefont {Chen}, \citenamefont {Draper}, \citenamefont {Gong},
  \citenamefont {Liu}, \citenamefont {Liu},\ and\ \citenamefont
  {Ma}}]{Yang:2014xsa}%
  \BibitemOpen
  \bibfield  {author} {\bibinfo {author} {\bibfnamefont {Yi-Bo}\ \bibnamefont
  {Yang}}, \bibinfo {author} {\bibfnamefont {Ying}\ \bibnamefont {Chen}},
  \bibinfo {author} {\bibfnamefont {Terrence}\ \bibnamefont {Draper}}, \bibinfo
  {author} {\bibfnamefont {Ming}\ \bibnamefont {Gong}}, \bibinfo {author}
  {\bibfnamefont {Keh-Fei}\ \bibnamefont {Liu}}, \bibinfo {author}
  {\bibfnamefont {Zhaofeng}\ \bibnamefont {Liu}}, \ and\ \bibinfo {author}
  {\bibfnamefont {Jian-Ping}\ \bibnamefont {Ma}},\ }\bibfield  {title}
  {\enquote {\bibinfo {title} {{Meson Mass Decomposition from Lattice QCD}},}\
  }\href {\doibase 10.1103/PhysRevD.91.074516} {\bibfield  {journal} {\bibinfo
  {journal} {Phys. Rev. D}\ }\textbf {\bibinfo {volume} {91}},\ \bibinfo
  {pages} {074516} (\bibinfo {year} {2015})},\ \Eprint
  {http://arxiv.org/abs/1405.4440} {arXiv:1405.4440 [hep-ph]} \BibitemShut
  {NoStop}%
\bibitem [{\citenamefont {L\"offler}\ \emph {et~al.}(2022)\citenamefont
  {L\"offler}, \citenamefont {Wein}, \citenamefont {Wurm}, \citenamefont
  {Weish\"aupl}, \citenamefont {Jenkins}, \citenamefont {R\"odl}, \citenamefont
  {Sch\"afer},\ and\ \citenamefont {Walter}}]{Loffler:2021afv}%
  \BibitemOpen
  \bibfield  {author} {\bibinfo {author} {\bibfnamefont {Marius}\ \bibnamefont
  {L\"offler}}, \bibinfo {author} {\bibfnamefont {Philipp}\ \bibnamefont
  {Wein}}, \bibinfo {author} {\bibfnamefont {Thomas}\ \bibnamefont {Wurm}},
  \bibinfo {author} {\bibfnamefont {Simon}\ \bibnamefont {Weish\"aupl}},
  \bibinfo {author} {\bibfnamefont {Daniel}\ \bibnamefont {Jenkins}}, \bibinfo
  {author} {\bibfnamefont {Rudolf}\ \bibnamefont {R\"odl}}, \bibinfo {author}
  {\bibfnamefont {Andreas}\ \bibnamefont {Sch\"afer}}, \ and\ \bibinfo {author}
  {\bibfnamefont {Lisa}\ \bibnamefont {Walter}} (\bibinfo {collaboration}
  {RQCD}),\ }\bibfield  {title} {\enquote {\bibinfo {title} {{Mellin moments of
  spin dependent and independent PDFs of the pion and rho meson}},}\ }\href
  {\doibase 10.1103/PhysRevD.105.014505} {\bibfield  {journal} {\bibinfo
  {journal} {Phys. Rev. D}\ }\textbf {\bibinfo {volume} {105}},\ \bibinfo
  {pages} {014505} (\bibinfo {year} {2022})},\ \Eprint
  {http://arxiv.org/abs/2108.07544} {arXiv:2108.07544 [hep-lat]} \BibitemShut
  {NoStop}%
\bibitem [{\citenamefont {Hackett}\ \emph {et~al.}(2023)\citenamefont
  {Hackett}, \citenamefont {Oare}, \citenamefont {Pefkou},\ and\ \citenamefont
  {Shanahan}}]{Hackett:2023nkr}%
  \BibitemOpen
  \bibfield  {author} {\bibinfo {author} {\bibfnamefont {Daniel~C.}\
  \bibnamefont {Hackett}}, \bibinfo {author} {\bibfnamefont {Patrick~R.}\
  \bibnamefont {Oare}}, \bibinfo {author} {\bibfnamefont {Dimitra~A.}\
  \bibnamefont {Pefkou}}, \ and\ \bibinfo {author} {\bibfnamefont {Phiala~E.}\
  \bibnamefont {Shanahan}},\ }\bibfield  {title} {\enquote {\bibinfo {title}
  {{Gravitational form factors of the pion from lattice QCD}},}\ }\href
  {\doibase 10.1103/PhysRevD.108.114504} {\bibfield  {journal} {\bibinfo
  {journal} {Phys. Rev. D}\ }\textbf {\bibinfo {volume} {108}},\ \bibinfo
  {pages} {114504} (\bibinfo {year} {2023})},\ \Eprint
  {http://arxiv.org/abs/2307.11707} {arXiv:2307.11707 [hep-lat]} \BibitemShut
  {NoStop}%
\bibitem [{\citenamefont {Sakai}\ and\ \citenamefont
  {Sugimoto}(2005{\natexlab{a}})}]{Sakai:2004cn}%
  \BibitemOpen
  \bibfield  {author} {\bibinfo {author} {\bibfnamefont {Tadakatsu}\
  \bibnamefont {Sakai}}\ and\ \bibinfo {author} {\bibfnamefont {Shigeki}\
  \bibnamefont {Sugimoto}},\ }\bibfield  {title} {\enquote {\bibinfo {title}
  {{Low energy hadron physics in holographic QCD}},}\ }\href {\doibase
  10.1143/PTP.113.843} {\bibfield  {journal} {\bibinfo  {journal} {Prog. Theor.
  Phys.}\ }\textbf {\bibinfo {volume} {113}},\ \bibinfo {pages} {843--882}
  (\bibinfo {year} {2005}{\natexlab{a}})},\ \Eprint
  {http://arxiv.org/abs/hep-th/0412141} {arXiv:hep-th/0412141} \BibitemShut
  {NoStop}%
\bibitem [{\citenamefont {Sakai}\ and\ \citenamefont
  {Sugimoto}(2005{\natexlab{b}})}]{Sakai:2005yt}%
  \BibitemOpen
  \bibfield  {author} {\bibinfo {author} {\bibfnamefont {Tadakatsu}\
  \bibnamefont {Sakai}}\ and\ \bibinfo {author} {\bibfnamefont {Shigeki}\
  \bibnamefont {Sugimoto}},\ }\bibfield  {title} {\enquote {\bibinfo {title}
  {{More on a holographic dual of QCD}},}\ }\href {\doibase
  10.1143/PTP.114.1083} {\bibfield  {journal} {\bibinfo  {journal} {Prog.
  Theor. Phys.}\ }\textbf {\bibinfo {volume} {114}},\ \bibinfo {pages}
  {1083--1118} (\bibinfo {year} {2005}{\natexlab{b}})},\ \Eprint
  {http://arxiv.org/abs/hep-th/0507073} {arXiv:hep-th/0507073} \BibitemShut
  {NoStop}%
\bibitem [{\citenamefont {Constable}\ and\ \citenamefont
  {Myers}(1999)}]{Constable:1999gb}%
  \BibitemOpen
  \bibfield  {author} {\bibinfo {author} {\bibfnamefont {Neil~R.}\ \bibnamefont
  {Constable}}\ and\ \bibinfo {author} {\bibfnamefont {Robert~C.}\ \bibnamefont
  {Myers}},\ }\bibfield  {title} {\enquote {\bibinfo {title} {{Spin two
  glueballs, positive energy theorems and the AdS / CFT correspondence}},}\
  }\href {\doibase 10.1088/1126-6708/1999/10/037} {\bibfield  {journal}
  {\bibinfo  {journal} {JHEP}\ }\textbf {\bibinfo {volume} {10}},\ \bibinfo
  {pages} {037} (\bibinfo {year} {1999})},\ \Eprint
  {http://arxiv.org/abs/hep-th/9908175} {arXiv:hep-th/9908175} \BibitemShut
  {NoStop}%
\bibitem [{\citenamefont {Brower}\ \emph {et~al.}(2000)\citenamefont {Brower},
  \citenamefont {Mathur},\ and\ \citenamefont {Tan}}]{Brower:2000rp}%
  \BibitemOpen
  \bibfield  {author} {\bibinfo {author} {\bibfnamefont {Richard~C.}\
  \bibnamefont {Brower}}, \bibinfo {author} {\bibfnamefont {Samir~D.}\
  \bibnamefont {Mathur}}, \ and\ \bibinfo {author} {\bibfnamefont {Chung-I}\
  \bibnamefont {Tan}},\ }\bibfield  {title} {\enquote {\bibinfo {title}
  {{Glueball spectrum for QCD from AdS supergravity duality}},}\ }\href
  {\doibase 10.1016/S0550-3213(00)00435-1} {\bibfield  {journal} {\bibinfo
  {journal} {Nucl. Phys. B}\ }\textbf {\bibinfo {volume} {587}},\ \bibinfo
  {pages} {249--276} (\bibinfo {year} {2000})},\ \Eprint
  {http://arxiv.org/abs/hep-th/0003115} {arXiv:hep-th/0003115} \BibitemShut
  {NoStop}%
\bibitem [{\citenamefont {Hata}\ \emph {et~al.}(2007)\citenamefont {Hata},
  \citenamefont {Sakai}, \citenamefont {Sugimoto},\ and\ \citenamefont
  {Yamato}}]{Hata:2007mb}%
  \BibitemOpen
  \bibfield  {author} {\bibinfo {author} {\bibfnamefont {Hiroyuki}\
  \bibnamefont {Hata}}, \bibinfo {author} {\bibfnamefont {Tadakatsu}\
  \bibnamefont {Sakai}}, \bibinfo {author} {\bibfnamefont {Shigeki}\
  \bibnamefont {Sugimoto}}, \ and\ \bibinfo {author} {\bibfnamefont
  {Shinichiro}\ \bibnamefont {Yamato}},\ }\bibfield  {title} {\enquote
  {\bibinfo {title} {{Baryons from instantons in holographic QCD}},}\ }\href
  {\doibase 10.1143/PTP.117.1157} {\bibfield  {journal} {\bibinfo  {journal}
  {Prog. Theor. Phys.}\ }\textbf {\bibinfo {volume} {117}},\ \bibinfo {pages}
  {1157} (\bibinfo {year} {2007})},\ \Eprint
  {http://arxiv.org/abs/hep-th/0701280} {arXiv:hep-th/0701280} \BibitemShut
  {NoStop}%
\bibitem [{\citenamefont {Hashimoto}\ \emph
  {et~al.}(2010{\natexlab{a}})\citenamefont {Hashimoto}, \citenamefont
  {Hirayama},\ and\ \citenamefont {Hong}}]{Hashimoto:2009hj}%
  \BibitemOpen
  \bibfield  {author} {\bibinfo {author} {\bibfnamefont {Koji}\ \bibnamefont
  {Hashimoto}}, \bibinfo {author} {\bibfnamefont {Takayuki}\ \bibnamefont
  {Hirayama}}, \ and\ \bibinfo {author} {\bibfnamefont {Deog~Ki}\ \bibnamefont
  {Hong}},\ }\bibfield  {title} {\enquote {\bibinfo {title} {{Quark Mass
  Dependence of Hadron Spectrum in Holographic QCD}},}\ }\href {\doibase
  10.1103/PhysRevD.81.045016} {\bibfield  {journal} {\bibinfo  {journal} {Phys.
  Rev. D}\ }\textbf {\bibinfo {volume} {81}},\ \bibinfo {pages} {045016}
  (\bibinfo {year} {2010}{\natexlab{a}})},\ \Eprint
  {http://arxiv.org/abs/0906.0402} {arXiv:0906.0402 [hep-th]} \BibitemShut
  {NoStop}%
\bibitem [{\citenamefont {Hashimoto}\ \emph
  {et~al.}(2010{\natexlab{b}})\citenamefont {Hashimoto}, \citenamefont
  {Iizuka}, \citenamefont {Ishii},\ and\ \citenamefont
  {Kadoh}}]{Hashimoto:2009st}%
  \BibitemOpen
  \bibfield  {author} {\bibinfo {author} {\bibfnamefont {Koji}\ \bibnamefont
  {Hashimoto}}, \bibinfo {author} {\bibfnamefont {Norihiro}\ \bibnamefont
  {Iizuka}}, \bibinfo {author} {\bibfnamefont {Takaaki}\ \bibnamefont {Ishii}},
  \ and\ \bibinfo {author} {\bibfnamefont {Daisuke}\ \bibnamefont {Kadoh}},\
  }\bibfield  {title} {\enquote {\bibinfo {title} {{Three-flavor quark mass
  dependence of baryon spectra in holographic QCD}},}\ }\href {\doibase
  10.1016/j.physletb.2010.06.008} {\bibfield  {journal} {\bibinfo  {journal}
  {Phys. Lett. B}\ }\textbf {\bibinfo {volume} {691}},\ \bibinfo {pages}
  {65--71} (\bibinfo {year} {2010}{\natexlab{b}})},\ \Eprint
  {http://arxiv.org/abs/0910.1179} {arXiv:0910.1179 [hep-th]} \BibitemShut
  {NoStop}%
\bibitem [{\citenamefont {Imoto}\ \emph {et~al.}(2010)\citenamefont {Imoto},
  \citenamefont {Sakai},\ and\ \citenamefont {Sugimoto}}]{Imoto:2010ef}%
  \BibitemOpen
  \bibfield  {author} {\bibinfo {author} {\bibfnamefont {Toshiya}\ \bibnamefont
  {Imoto}}, \bibinfo {author} {\bibfnamefont {Tadakatsu}\ \bibnamefont
  {Sakai}}, \ and\ \bibinfo {author} {\bibfnamefont {Shigeki}\ \bibnamefont
  {Sugimoto}},\ }\bibfield  {title} {\enquote {\bibinfo {title} {{Mesons as
  Open Strings in a Holographic Dual of QCD}},}\ }\href {\doibase
  10.1143/PTP.124.263} {\bibfield  {journal} {\bibinfo  {journal} {Prog. Theor.
  Phys.}\ }\textbf {\bibinfo {volume} {124}},\ \bibinfo {pages} {263--284}
  (\bibinfo {year} {2010})},\ \Eprint {http://arxiv.org/abs/1005.0655}
  {arXiv:1005.0655 [hep-th]} \BibitemShut {NoStop}%
\bibitem [{\citenamefont {Ishii}(2011)}]{Ishii:2010ib}%
  \BibitemOpen
  \bibfield  {author} {\bibinfo {author} {\bibfnamefont {Takaaki}\ \bibnamefont
  {Ishii}},\ }\bibfield  {title} {\enquote {\bibinfo {title} {{Toward
  Bound-State Approach to Strangeness in Holographic QCD}},}\ }\href {\doibase
  10.1016/j.physletb.2010.11.043} {\bibfield  {journal} {\bibinfo  {journal}
  {Phys. Lett. B}\ }\textbf {\bibinfo {volume} {695}},\ \bibinfo {pages}
  {392--396} (\bibinfo {year} {2011})},\ \Eprint
  {http://arxiv.org/abs/1009.0986} {arXiv:1009.0986 [hep-th]} \BibitemShut
  {NoStop}%
\bibitem [{\citenamefont {Liu}\ and\ \citenamefont
  {Zahed}(2017{\natexlab{a}})}]{Liu:2017xzo}%
  \BibitemOpen
  \bibfield  {author} {\bibinfo {author} {\bibfnamefont {Yizhuang}\
  \bibnamefont {Liu}}\ and\ \bibinfo {author} {\bibfnamefont {Ismail}\
  \bibnamefont {Zahed}},\ }\bibfield  {title} {\enquote {\bibinfo {title}
  {{Heavy Baryons and their Exotics from Instantons in Holographic QCD}},}\
  }\href {\doibase 10.1103/PhysRevD.95.116012} {\bibfield  {journal} {\bibinfo
  {journal} {Phys. Rev. D}\ }\textbf {\bibinfo {volume} {95}},\ \bibinfo
  {pages} {116012} (\bibinfo {year} {2017}{\natexlab{a}})},\ \Eprint
  {http://arxiv.org/abs/1704.03412} {arXiv:1704.03412 [hep-ph]} \BibitemShut
  {NoStop}%
\bibitem [{\citenamefont {Liu}\ and\ \citenamefont
  {Zahed}(2017{\natexlab{b}})}]{Liu:2017frj}%
  \BibitemOpen
  \bibfield  {author} {\bibinfo {author} {\bibfnamefont {Yizhuang}\
  \bibnamefont {Liu}}\ and\ \bibinfo {author} {\bibfnamefont {Ismail}\
  \bibnamefont {Zahed}},\ }\bibfield  {title} {\enquote {\bibinfo {title}
  {{Heavy and Strange Holographic Baryons}},}\ }\href {\doibase
  10.1103/PhysRevD.96.056027} {\bibfield  {journal} {\bibinfo  {journal} {Phys.
  Rev. D}\ }\textbf {\bibinfo {volume} {96}},\ \bibinfo {pages} {056027}
  (\bibinfo {year} {2017}{\natexlab{b}})},\ \Eprint
  {http://arxiv.org/abs/1705.01397} {arXiv:1705.01397 [hep-ph]} \BibitemShut
  {NoStop}%
\bibitem [{\citenamefont {Hashimoto}\ \emph {et~al.}(2019)\citenamefont
  {Hashimoto}, \citenamefont {Matsuo},\ and\ \citenamefont
  {Morita}}]{Hashimoto:2019wmg}%
  \BibitemOpen
  \bibfield  {author} {\bibinfo {author} {\bibfnamefont {Koji}\ \bibnamefont
  {Hashimoto}}, \bibinfo {author} {\bibfnamefont {Yoshinori}\ \bibnamefont
  {Matsuo}}, \ and\ \bibinfo {author} {\bibfnamefont {Takeshi}\ \bibnamefont
  {Morita}},\ }\bibfield  {title} {\enquote {\bibinfo {title} {{Nuclear states
  and spectra in holographic QCD}},}\ }\href {\doibase 10.1007/JHEP12(2019)001}
  {\bibfield  {journal} {\bibinfo  {journal} {JHEP}\ }\textbf {\bibinfo
  {volume} {12}},\ \bibinfo {pages} {001} (\bibinfo {year} {2019})},\ \Eprint
  {http://arxiv.org/abs/1902.07444} {arXiv:1902.07444 [hep-th]} \BibitemShut
  {NoStop}%
\bibitem [{\citenamefont {Liu}\ \emph {et~al.}(2022{\natexlab{a}})\citenamefont
  {Liu}, \citenamefont {Nowak},\ and\ \citenamefont {Zahed}}]{Liu:2019yye}%
  \BibitemOpen
  \bibfield  {author} {\bibinfo {author} {\bibfnamefont {Yizhuang}\
  \bibnamefont {Liu}}, \bibinfo {author} {\bibfnamefont {Maciej~A.}\
  \bibnamefont {Nowak}}, \ and\ \bibinfo {author} {\bibfnamefont {Ismail}\
  \bibnamefont {Zahed}},\ }\bibfield  {title} {\enquote {\bibinfo {title}
  {{Holographic tetraquarks and the newly observed Tcc+ at LHCb}},}\ }\href
  {\doibase 10.1103/PhysRevD.105.054021} {\bibfield  {journal} {\bibinfo
  {journal} {Phys. Rev. D}\ }\textbf {\bibinfo {volume} {105}},\ \bibinfo
  {pages} {054021} (\bibinfo {year} {2022}{\natexlab{a}})},\ \Eprint
  {http://arxiv.org/abs/1909.02497} {arXiv:1909.02497 [hep-ph]} \BibitemShut
  {NoStop}%
\bibitem [{\citenamefont {Nakas}\ and\ \citenamefont
  {Rigatos}(2020)}]{Nakas:2020hyo}%
  \BibitemOpen
  \bibfield  {author} {\bibinfo {author} {\bibfnamefont {Theodoros}\
  \bibnamefont {Nakas}}\ and\ \bibinfo {author} {\bibfnamefont
  {Konstantinos~S.}\ \bibnamefont {Rigatos}},\ }\bibfield  {title} {\enquote
  {\bibinfo {title} {{Fermions and baryons as open-string states from brane
  junctions}},}\ }\href {\doibase 10.1007/JHEP12(2020)157} {\bibfield
  {journal} {\bibinfo  {journal} {JHEP}\ }\textbf {\bibinfo {volume} {12}},\
  \bibinfo {pages} {157} (\bibinfo {year} {2020})},\ \Eprint
  {http://arxiv.org/abs/2010.00025} {arXiv:2010.00025 [hep-th]} \BibitemShut
  {NoStop}%
\bibitem [{\citenamefont {Hayashi}\ \emph {et~al.}(2020)\citenamefont
  {Hayashi}, \citenamefont {Ogino}, \citenamefont {Sakai},\ and\ \citenamefont
  {Sugimoto}}]{Hayashi:2020ipd}%
  \BibitemOpen
  \bibfield  {author} {\bibinfo {author} {\bibfnamefont {Yasuhiro}\
  \bibnamefont {Hayashi}}, \bibinfo {author} {\bibfnamefont {Takahiro}\
  \bibnamefont {Ogino}}, \bibinfo {author} {\bibfnamefont {Tadakatsu}\
  \bibnamefont {Sakai}}, \ and\ \bibinfo {author} {\bibfnamefont {Shigeki}\
  \bibnamefont {Sugimoto}},\ }\bibfield  {title} {\enquote {\bibinfo {title}
  {{Stringy excited baryons in holographic quantum chromodynamics}},}\ }\href
  {\doibase 10.1093/ptep/ptaa045} {\bibfield  {journal} {\bibinfo  {journal}
  {PTEP}\ }\textbf {\bibinfo {volume} {2020}},\ \bibinfo {pages} {053B04}
  (\bibinfo {year} {2020})},\ \Eprint {http://arxiv.org/abs/2001.01461}
  {arXiv:2001.01461 [hep-th]} \BibitemShut {NoStop}%
\bibitem [{\citenamefont {Fujii}\ and\ \citenamefont
  {Hosaka}(2020)}]{Fujii:2020jre}%
  \BibitemOpen
  \bibfield  {author} {\bibinfo {author} {\bibfnamefont {Daisuke}\ \bibnamefont
  {Fujii}}\ and\ \bibinfo {author} {\bibfnamefont {Atsushi}\ \bibnamefont
  {Hosaka}},\ }\bibfield  {title} {\enquote {\bibinfo {title} {{Heavy baryons
  in holographic QCD with higher dimensional degrees of freedom}},}\ }\href
  {\doibase 10.1103/PhysRevD.101.126008} {\bibfield  {journal} {\bibinfo
  {journal} {Phys. Rev. D}\ }\textbf {\bibinfo {volume} {101}},\ \bibinfo
  {pages} {126008} (\bibinfo {year} {2020})},\ \Eprint
  {http://arxiv.org/abs/2003.13415} {arXiv:2003.13415 [hep-ph]} \BibitemShut
  {NoStop}%
\bibitem [{\citenamefont {Suganuma}\ and\ \citenamefont
  {Hori}(2020)}]{Suganuma:2020jng}%
  \BibitemOpen
  \bibfield  {author} {\bibinfo {author} {\bibfnamefont {Hideo}\ \bibnamefont
  {Suganuma}}\ and\ \bibinfo {author} {\bibfnamefont {Keiichiro}\ \bibnamefont
  {Hori}},\ }\bibfield  {title} {\enquote {\bibinfo {title} {{Topological
  Objects in Holographic QCD}},}\ }\href {\doibase 10.1088/1402-4896/ab986c}
  {\bibfield  {journal} {\bibinfo  {journal} {Phys. Scripta}\ }\textbf
  {\bibinfo {volume} {95}},\ \bibinfo {pages} {074014} (\bibinfo {year}
  {2020})},\ \Eprint {http://arxiv.org/abs/2003.07127} {arXiv:2003.07127
  [hep-th]} \BibitemShut {NoStop}%
\bibitem [{\citenamefont {Liu}\ \emph {et~al.}(2021{\natexlab{a}})\citenamefont
  {Liu}, \citenamefont {Nowak},\ and\ \citenamefont {Zahed}}]{Liu:2021tpq}%
  \BibitemOpen
  \bibfield  {author} {\bibinfo {author} {\bibfnamefont {Yizhuang}\
  \bibnamefont {Liu}}, \bibinfo {author} {\bibfnamefont {Maciej~A.}\
  \bibnamefont {Nowak}}, \ and\ \bibinfo {author} {\bibfnamefont {Ismail}\
  \bibnamefont {Zahed}},\ }\bibfield  {title} {\enquote {\bibinfo {title}
  {{Holographic charm and bottom pentaquarks. I. Mass spectra with spin
  effects}},}\ }\href {\doibase 10.1103/PhysRevD.104.114021} {\bibfield
  {journal} {\bibinfo  {journal} {Phys. Rev. D}\ }\textbf {\bibinfo {volume}
  {104}},\ \bibinfo {pages} {114021} (\bibinfo {year} {2021}{\natexlab{a}})},\
  \Eprint {http://arxiv.org/abs/2108.04334} {arXiv:2108.04334 [hep-ph]}
  \BibitemShut {NoStop}%
\bibitem [{\citenamefont {Liu}\ \emph {et~al.}(2022{\natexlab{b}})\citenamefont
  {Liu}, \citenamefont {Nowak},\ and\ \citenamefont {Zahed}}]{Liu:2022urb}%
  \BibitemOpen
  \bibfield  {author} {\bibinfo {author} {\bibfnamefont {Yizhuang}\
  \bibnamefont {Liu}}, \bibinfo {author} {\bibfnamefont {Maciej~A.}\
  \bibnamefont {Nowak}}, \ and\ \bibinfo {author} {\bibfnamefont {Ismail}\
  \bibnamefont {Zahed}},\ }\bibfield  {title} {\enquote {\bibinfo {title}
  {{Hyperons and \ensuremath{\Theta}s+ in holographic QCD}},}\ }\href {\doibase
  10.1103/PhysRevD.105.114021} {\bibfield  {journal} {\bibinfo  {journal}
  {Phys. Rev. D}\ }\textbf {\bibinfo {volume} {105}},\ \bibinfo {pages}
  {114021} (\bibinfo {year} {2022}{\natexlab{b}})},\ \Eprint
  {http://arxiv.org/abs/2201.01791} {arXiv:2201.01791 [hep-ph]} \BibitemShut
  {NoStop}%
\bibitem [{\citenamefont {Hashimoto}\ \emph
  {et~al.}(2008{\natexlab{a}})\citenamefont {Hashimoto}, \citenamefont {Tan},\
  and\ \citenamefont {Terashima}}]{Hashimoto:2007ze}%
  \BibitemOpen
  \bibfield  {author} {\bibinfo {author} {\bibfnamefont {Koji}\ \bibnamefont
  {Hashimoto}}, \bibinfo {author} {\bibfnamefont {Chung-I}\ \bibnamefont
  {Tan}}, \ and\ \bibinfo {author} {\bibfnamefont {Seiji}\ \bibnamefont
  {Terashima}},\ }\bibfield  {title} {\enquote {\bibinfo {title} {{Glueball
  decay in holographic QCD}},}\ }\href {\doibase 10.1103/PhysRevD.77.086001}
  {\bibfield  {journal} {\bibinfo  {journal} {Phys. Rev. D}\ }\textbf {\bibinfo
  {volume} {77}},\ \bibinfo {pages} {086001} (\bibinfo {year}
  {2008}{\natexlab{a}})},\ \Eprint {http://arxiv.org/abs/0709.2208}
  {arXiv:0709.2208 [hep-th]} \BibitemShut {NoStop}%
\bibitem [{\citenamefont {Hashimoto}\ \emph
  {et~al.}(2008{\natexlab{b}})\citenamefont {Hashimoto}, \citenamefont
  {Sakai},\ and\ \citenamefont {Sugimoto}}]{Hashimoto:2008zw}%
  \BibitemOpen
  \bibfield  {author} {\bibinfo {author} {\bibfnamefont {Koji}\ \bibnamefont
  {Hashimoto}}, \bibinfo {author} {\bibfnamefont {Tadakatsu}\ \bibnamefont
  {Sakai}}, \ and\ \bibinfo {author} {\bibfnamefont {Shigeki}\ \bibnamefont
  {Sugimoto}},\ }\bibfield  {title} {\enquote {\bibinfo {title} {{Holographic
  Baryons: Static Properties and Form Factors from Gauge/String Duality}},}\
  }\href {\doibase 10.1143/PTP.120.1093} {\bibfield  {journal} {\bibinfo
  {journal} {Prog. Theor. Phys.}\ }\textbf {\bibinfo {volume} {120}},\ \bibinfo
  {pages} {1093--1137} (\bibinfo {year} {2008}{\natexlab{b}})},\ \Eprint
  {http://arxiv.org/abs/0806.3122} {arXiv:0806.3122 [hep-th]} \BibitemShut
  {NoStop}%
\bibitem [{\citenamefont {Hata}\ \emph {et~al.}(2008)\citenamefont {Hata},
  \citenamefont {Murata},\ and\ \citenamefont {Yamato}}]{Hata:2008xc}%
  \BibitemOpen
  \bibfield  {author} {\bibinfo {author} {\bibfnamefont {Hiroyuki}\
  \bibnamefont {Hata}}, \bibinfo {author} {\bibfnamefont {Masaki}\ \bibnamefont
  {Murata}}, \ and\ \bibinfo {author} {\bibfnamefont {Shinichiro}\ \bibnamefont
  {Yamato}},\ }\bibfield  {title} {\enquote {\bibinfo {title} {{Chiral currents
  and static properties of nucleons in holographic QCD}},}\ }\href {\doibase
  10.1103/PhysRevD.78.086006} {\bibfield  {journal} {\bibinfo  {journal} {Phys.
  Rev. D}\ }\textbf {\bibinfo {volume} {78}},\ \bibinfo {pages} {086006}
  (\bibinfo {year} {2008})},\ \Eprint {http://arxiv.org/abs/0803.0180}
  {arXiv:0803.0180 [hep-th]} \BibitemShut {NoStop}%
\bibitem [{\citenamefont {Kim}\ and\ \citenamefont {Zahed}(2008)}]{Kim:2008pw}%
  \BibitemOpen
  \bibfield  {author} {\bibinfo {author} {\bibfnamefont {Keun-Young}\
  \bibnamefont {Kim}}\ and\ \bibinfo {author} {\bibfnamefont {Ismail}\
  \bibnamefont {Zahed}},\ }\bibfield  {title} {\enquote {\bibinfo {title}
  {{Electromagnetic Baryon Form Factors from Holographic QCD}},}\ }\href
  {\doibase 10.1088/1126-6708/2008/09/007} {\bibfield  {journal} {\bibinfo
  {journal} {JHEP}\ }\textbf {\bibinfo {volume} {09}},\ \bibinfo {pages} {007}
  (\bibinfo {year} {2008})},\ \Eprint {http://arxiv.org/abs/0807.0033}
  {arXiv:0807.0033 [hep-th]} \BibitemShut {NoStop}%
\bibitem [{\citenamefont {Grigoryan}\ \emph {et~al.}(2009)\citenamefont
  {Grigoryan}, \citenamefont {Lee},\ and\ \citenamefont
  {Yee}}]{Grigoryan:2009pp}%
  \BibitemOpen
  \bibfield  {author} {\bibinfo {author} {\bibfnamefont {Hovhannes~R.}\
  \bibnamefont {Grigoryan}}, \bibinfo {author} {\bibfnamefont {T.~S.~H.}\
  \bibnamefont {Lee}}, \ and\ \bibinfo {author} {\bibfnamefont {Ho-Ung}\
  \bibnamefont {Yee}},\ }\bibfield  {title} {\enquote {\bibinfo {title}
  {{Electromagnetic Nucleon-to-Delta Transition in Holographic QCD}},}\ }\href
  {\doibase 10.1103/PhysRevD.80.055006} {\bibfield  {journal} {\bibinfo
  {journal} {Phys. Rev. D}\ }\textbf {\bibinfo {volume} {80}},\ \bibinfo
  {pages} {055006} (\bibinfo {year} {2009})},\ \Eprint
  {http://arxiv.org/abs/0904.3710} {arXiv:0904.3710 [hep-ph]} \BibitemShut
  {NoStop}%
\bibitem [{\citenamefont {Ballon~Bayona}\ \emph
  {et~al.}(2010{\natexlab{a}})\citenamefont {Ballon~Bayona}, \citenamefont
  {Boschi-Filho}, \citenamefont {Braga},\ and\ \citenamefont
  {Torres}}]{BallonBayona:2009ar}%
  \BibitemOpen
  \bibfield  {author} {\bibinfo {author} {\bibfnamefont {C.~A.}\ \bibnamefont
  {Ballon~Bayona}}, \bibinfo {author} {\bibfnamefont {Henrique}\ \bibnamefont
  {Boschi-Filho}}, \bibinfo {author} {\bibfnamefont {Nelson R.~F.}\
  \bibnamefont {Braga}}, \ and\ \bibinfo {author} {\bibfnamefont {Marcus
  A.~C.}\ \bibnamefont {Torres}},\ }\bibfield  {title} {\enquote {\bibinfo
  {title} {{Form factors of vector and axial-vector mesons in holographic D4-D8
  model}},}\ }\href {\doibase 10.1007/JHEP01(2010)052} {\bibfield  {journal}
  {\bibinfo  {journal} {JHEP}\ }\textbf {\bibinfo {volume} {01}},\ \bibinfo
  {pages} {052} (\bibinfo {year} {2010}{\natexlab{a}})},\ \Eprint
  {http://arxiv.org/abs/0911.0023} {arXiv:0911.0023 [hep-th]} \BibitemShut
  {NoStop}%
\bibitem [{\citenamefont {Bayona}\ \emph {et~al.}(2010)\citenamefont {Bayona},
  \citenamefont {Boschi-Filho}, \citenamefont {Braga},\ and\ \citenamefont
  {Torres}}]{Bayona:2009pk}%
  \BibitemOpen
  \bibfield  {author} {\bibinfo {author} {\bibfnamefont {C.~A.~Ballon}\
  \bibnamefont {Bayona}}, \bibinfo {author} {\bibfnamefont {Henrique}\
  \bibnamefont {Boschi-Filho}}, \bibinfo {author} {\bibfnamefont {Nelson
  R.~F.}\ \bibnamefont {Braga}}, \ and\ \bibinfo {author} {\bibfnamefont
  {Marcus A.~C.}\ \bibnamefont {Torres}},\ }\bibfield  {title} {\enquote
  {\bibinfo {title} {{Scattering vector mesons in D4 / D8 model}},}\ }\href
  {\doibase 10.1016/j.nuclphysbps.2010.02.015} {\bibfield  {journal} {\bibinfo
  {journal} {Nucl. Phys. B Proc. Suppl.}\ }\textbf {\bibinfo {volume} {199}},\
  \bibinfo {pages} {119--124} (\bibinfo {year} {2010})},\ \Eprint
  {http://arxiv.org/abs/0912.0191} {arXiv:0912.0191 [hep-th]} \BibitemShut
  {NoStop}%
\bibitem [{\citenamefont {Ballon~Bayona}\ \emph
  {et~al.}(2010{\natexlab{b}})\citenamefont {Ballon~Bayona}, \citenamefont
  {Boschi-Filho}, \citenamefont {Braga},\ and\ \citenamefont
  {Torres}}]{BallonBayona:2010ae}%
  \BibitemOpen
  \bibfield  {author} {\bibinfo {author} {\bibfnamefont {C.~A.}\ \bibnamefont
  {Ballon~Bayona}}, \bibinfo {author} {\bibfnamefont {Henrique}\ \bibnamefont
  {Boschi-Filho}}, \bibinfo {author} {\bibfnamefont {Nelson R.~F.}\
  \bibnamefont {Braga}}, \ and\ \bibinfo {author} {\bibfnamefont {Marcus
  A.~C.}\ \bibnamefont {Torres}},\ }\bibfield  {title} {\enquote {\bibinfo
  {title} {{Deep inelastic scattering for vector mesons in holographic D4-D8
  model}},}\ }\href {\doibase 10.1007/JHEP10(2010)055} {\bibfield  {journal}
  {\bibinfo  {journal} {JHEP}\ }\textbf {\bibinfo {volume} {10}},\ \bibinfo
  {pages} {055} (\bibinfo {year} {2010}{\natexlab{b}})},\ \Eprint
  {http://arxiv.org/abs/1007.2448} {arXiv:1007.2448 [hep-th]} \BibitemShut
  {NoStop}%
\bibitem [{\citenamefont {Cherman}\ and\ \citenamefont
  {Ishii}(2012)}]{Cherman:2011ve}%
  \BibitemOpen
  \bibfield  {author} {\bibinfo {author} {\bibfnamefont {Aleksey}\ \bibnamefont
  {Cherman}}\ and\ \bibinfo {author} {\bibfnamefont {Takaaki}\ \bibnamefont
  {Ishii}},\ }\bibfield  {title} {\enquote {\bibinfo {title} {{Long-distance
  properties of baryons in the Sakai-Sugimoto model}},}\ }\href {\doibase
  10.1103/PhysRevD.86.045011} {\bibfield  {journal} {\bibinfo  {journal} {Phys.
  Rev. D}\ }\textbf {\bibinfo {volume} {86}},\ \bibinfo {pages} {045011}
  (\bibinfo {year} {2012})},\ \Eprint {http://arxiv.org/abs/1109.4665}
  {arXiv:1109.4665 [hep-th]} \BibitemShut {NoStop}%
\bibitem [{\citenamefont {Bayona}\ \emph {et~al.}(2013)\citenamefont {Bayona},
  \citenamefont {Boschi-Filho}, \citenamefont {Braga}, \citenamefont {Ihl},\
  and\ \citenamefont {Torres}}]{Bayona:2011xj}%
  \BibitemOpen
  \bibfield  {author} {\bibinfo {author} {\bibfnamefont {C.~A.~Ballon}\
  \bibnamefont {Bayona}}, \bibinfo {author} {\bibfnamefont {Henrique}\
  \bibnamefont {Boschi-Filho}}, \bibinfo {author} {\bibfnamefont {Nelson
  R.~F.}\ \bibnamefont {Braga}}, \bibinfo {author} {\bibfnamefont {Matthias}\
  \bibnamefont {Ihl}}, \ and\ \bibinfo {author} {\bibfnamefont {Marcus A.~C.}\
  \bibnamefont {Torres}},\ }\bibfield  {title} {\enquote {\bibinfo {title}
  {{Generalized baryon form factors and proton structure functions in the
  Sakai-Sugimoto model}},}\ }\href {\doibase 10.1016/j.nuclphysb.2012.08.017}
  {\bibfield  {journal} {\bibinfo  {journal} {Nucl. Phys. B}\ }\textbf
  {\bibinfo {volume} {866}},\ \bibinfo {pages} {124--156} (\bibinfo {year}
  {2013})},\ \Eprint {http://arxiv.org/abs/1112.1439} {arXiv:1112.1439
  [hep-ph]} \BibitemShut {NoStop}%
\bibitem [{\citenamefont {Harada}\ and\ \citenamefont
  {Rho}(2011)}]{Harada:2011ur}%
  \BibitemOpen
  \bibfield  {author} {\bibinfo {author} {\bibfnamefont {Masayasu}\
  \bibnamefont {Harada}}\ and\ \bibinfo {author} {\bibfnamefont {Mannque}\
  \bibnamefont {Rho}},\ }\bibfield  {title} {\enquote {\bibinfo {title}
  {{Integrating Holographic Vector Dominance to Hidden Local Symmetry for the
  Nucleon Form Factor}},}\ }\href {\doibase 10.1103/PhysRevD.83.114040}
  {\bibfield  {journal} {\bibinfo  {journal} {Phys. Rev. D}\ }\textbf {\bibinfo
  {volume} {83}},\ \bibinfo {pages} {114040} (\bibinfo {year} {2011})},\
  \Eprint {http://arxiv.org/abs/1102.5489} {arXiv:1102.5489 [hep-ph]}
  \BibitemShut {NoStop}%
\bibitem [{\citenamefont {Br\"unner}\ \emph {et~al.}(2015)\citenamefont
  {Br\"unner}, \citenamefont {Parganlija},\ and\ \citenamefont
  {Rebhan}}]{Brunner:2015oqa}%
  \BibitemOpen
  \bibfield  {author} {\bibinfo {author} {\bibfnamefont {Frederic}\
  \bibnamefont {Br\"unner}}, \bibinfo {author} {\bibfnamefont {Denis}\
  \bibnamefont {Parganlija}}, \ and\ \bibinfo {author} {\bibfnamefont {Anton}\
  \bibnamefont {Rebhan}},\ }\bibfield  {title} {\enquote {\bibinfo {title}
  {{Glueball Decay Rates in the Witten-Sakai-Sugimoto Model}},}\ }\href
  {\doibase 10.1103/PhysRevD.91.106002} {\bibfield  {journal} {\bibinfo
  {journal} {Phys. Rev. D}\ }\textbf {\bibinfo {volume} {91}},\ \bibinfo
  {pages} {106002} (\bibinfo {year} {2015})},\ \bibinfo {note} {[Erratum:
  Phys.Rev.D 93, 109903 (2016)]},\ \Eprint {http://arxiv.org/abs/1501.07906}
  {arXiv:1501.07906 [hep-ph]} \BibitemShut {NoStop}%
\bibitem [{\citenamefont {Li}(2017)}]{Li:2015oza}%
  \BibitemOpen
  \bibfield  {author} {\bibinfo {author} {\bibfnamefont {Si-wen}\ \bibnamefont
  {Li}},\ }\bibfield  {title} {\enquote {\bibinfo {title}
  {{Glueball\textendash{}baryon interactions in holographic QCD}},}\ }\href
  {\doibase 10.1016/j.physletb.2017.08.011} {\bibfield  {journal} {\bibinfo
  {journal} {Phys. Lett. B}\ }\textbf {\bibinfo {volume} {773}},\ \bibinfo
  {pages} {142--149} (\bibinfo {year} {2017})},\ \Eprint
  {http://arxiv.org/abs/1509.06914} {arXiv:1509.06914 [hep-th]} \BibitemShut
  {NoStop}%
\bibitem [{\citenamefont {Druks}\ \emph {et~al.}(2019)\citenamefont {Druks},
  \citenamefont {Lau},\ and\ \citenamefont {Zahed}}]{Druks:2018hif}%
  \BibitemOpen
  \bibfield  {author} {\bibinfo {author} {\bibfnamefont {Ori~C.}\ \bibnamefont
  {Druks}}, \bibinfo {author} {\bibfnamefont {Pang H.~C.}\ \bibnamefont {Lau}},
  \ and\ \bibinfo {author} {\bibfnamefont {Ismail}\ \bibnamefont {Zahed}},\
  }\bibfield  {title} {\enquote {\bibinfo {title} {{Electromagnetic and Axial
  Current Form Factors and Spectroscopy of Three-Flavor Holographic
  Baryons}},}\ }\href {\doibase 10.1103/PhysRevD.99.054022} {\bibfield
  {journal} {\bibinfo  {journal} {Phys. Rev. D}\ }\textbf {\bibinfo {volume}
  {99}},\ \bibinfo {pages} {054022} (\bibinfo {year} {2019})},\ \Eprint
  {http://arxiv.org/abs/1807.05956} {arXiv:1807.05956 [hep-ph]} \BibitemShut
  {NoStop}%
\bibitem [{\citenamefont {Fujii}\ and\ \citenamefont
  {Hosaka}(2021)}]{Fujii:2021tsw}%
  \BibitemOpen
  \bibfield  {author} {\bibinfo {author} {\bibfnamefont {Daisuke}\ \bibnamefont
  {Fujii}}\ and\ \bibinfo {author} {\bibfnamefont {Atsushi}\ \bibnamefont
  {Hosaka}},\ }\bibfield  {title} {\enquote {\bibinfo {title} {{Decay
  properties of Roper resonance in the holographic QCD}},}\ }\href {\doibase
  10.1103/PhysRevD.104.014022} {\bibfield  {journal} {\bibinfo  {journal}
  {Phys. Rev. D}\ }\textbf {\bibinfo {volume} {104}},\ \bibinfo {pages}
  {014022} (\bibinfo {year} {2021})},\ \Eprint
  {http://arxiv.org/abs/2104.02227} {arXiv:2104.02227 [hep-ph]} \BibitemShut
  {NoStop}%
\bibitem [{\citenamefont {Liu}\ \emph {et~al.}(2021{\natexlab{b}})\citenamefont
  {Liu}, \citenamefont {Nowak},\ and\ \citenamefont {Zahed}}]{Liu:2021ixf}%
  \BibitemOpen
  \bibfield  {author} {\bibinfo {author} {\bibfnamefont {Yizhuang}\
  \bibnamefont {Liu}}, \bibinfo {author} {\bibfnamefont {Maciej~A.}\
  \bibnamefont {Nowak}}, \ and\ \bibinfo {author} {\bibfnamefont {Ismail}\
  \bibnamefont {Zahed}},\ }\bibfield  {title} {\enquote {\bibinfo {title}
  {{Holographic charm and bottom pentaquarks. II. Open and hidden decay
  widths}},}\ }\href {\doibase 10.1103/PhysRevD.104.114022} {\bibfield
  {journal} {\bibinfo  {journal} {Phys. Rev. D}\ }\textbf {\bibinfo {volume}
  {104}},\ \bibinfo {pages} {114022} (\bibinfo {year} {2021}{\natexlab{b}})},\
  \Eprint {http://arxiv.org/abs/2108.07074} {arXiv:2108.07074 [hep-ph]}
  \BibitemShut {NoStop}%
\bibitem [{\citenamefont {Liu}\ \emph {et~al.}(2021{\natexlab{c}})\citenamefont
  {Liu}, \citenamefont {Mamo}, \citenamefont {Nowak},\ and\ \citenamefont
  {Zahed}}]{Liu:2021efc}%
  \BibitemOpen
  \bibfield  {author} {\bibinfo {author} {\bibfnamefont {Yizhuang}\
  \bibnamefont {Liu}}, \bibinfo {author} {\bibfnamefont {Kiminad~A.}\
  \bibnamefont {Mamo}}, \bibinfo {author} {\bibfnamefont {Maciej~A.}\
  \bibnamefont {Nowak}}, \ and\ \bibinfo {author} {\bibfnamefont {Ismail}\
  \bibnamefont {Zahed}},\ }\bibfield  {title} {\enquote {\bibinfo {title}
  {{Holographic charm and bottom pentaquarks. III. Excitations through
  photoproduction of heavy mesons}},}\ }\href {\doibase
  10.1103/PhysRevD.104.114023} {\bibfield  {journal} {\bibinfo  {journal}
  {Phys. Rev. D}\ }\textbf {\bibinfo {volume} {104}},\ \bibinfo {pages}
  {114023} (\bibinfo {year} {2021}{\natexlab{c}})},\ \Eprint
  {http://arxiv.org/abs/2109.03103} {arXiv:2109.03103 [hep-ph]} \BibitemShut
  {NoStop}%
\bibitem [{\citenamefont {Iwanaka}\ \emph {et~al.}(2022)\citenamefont
  {Iwanaka}, \citenamefont {Fujii},\ and\ \citenamefont
  {Hosaka}}]{Iwanaka:2022uje}%
  \BibitemOpen
  \bibfield  {author} {\bibinfo {author} {\bibfnamefont {Akihiro}\ \bibnamefont
  {Iwanaka}}, \bibinfo {author} {\bibfnamefont {Daisuke}\ \bibnamefont
  {Fujii}}, \ and\ \bibinfo {author} {\bibfnamefont {Atsushi}\ \bibnamefont
  {Hosaka}},\ }\bibfield  {title} {\enquote {\bibinfo {title} {{Decay
  properties of N(1535) in the holographic QCD}},}\ }\href {\doibase
  10.1103/PhysRevD.105.114057} {\bibfield  {journal} {\bibinfo  {journal}
  {Phys. Rev. D}\ }\textbf {\bibinfo {volume} {105}},\ \bibinfo {pages}
  {114057} (\bibinfo {year} {2022})},\ \Eprint
  {http://arxiv.org/abs/2205.06054} {arXiv:2205.06054 [hep-ph]} \BibitemShut
  {NoStop}%
\bibitem [{\citenamefont {Fujii}\ \emph {et~al.}(2022)\citenamefont {Fujii},
  \citenamefont {Iwanaka},\ and\ \citenamefont {Hosaka}}]{Fujii:2022yqh}%
  \BibitemOpen
  \bibfield  {author} {\bibinfo {author} {\bibfnamefont {Daisuke}\ \bibnamefont
  {Fujii}}, \bibinfo {author} {\bibfnamefont {Akihiro}\ \bibnamefont
  {Iwanaka}}, \ and\ \bibinfo {author} {\bibfnamefont {Atsushi}\ \bibnamefont
  {Hosaka}},\ }\bibfield  {title} {\enquote {\bibinfo {title} {{Electromagnetic
  transition amplitude for Roper resonance from holographic QCD}},}\ }\href
  {\doibase 10.1103/PhysRevD.106.014010} {\bibfield  {journal} {\bibinfo
  {journal} {Phys. Rev. D}\ }\textbf {\bibinfo {volume} {106}},\ \bibinfo
  {pages} {014010} (\bibinfo {year} {2022})},\ \Eprint
  {http://arxiv.org/abs/2203.13988} {arXiv:2203.13988 [hep-ph]} \BibitemShut
  {NoStop}%
\bibitem [{\citenamefont {Fujii}(2023)}]{Fujii:2023ajs}%
  \BibitemOpen
  \bibfield  {author} {\bibinfo {author} {\bibfnamefont {Daisuke}\ \bibnamefont
  {Fujii}},\ }\emph {\bibinfo {title} {{Dynamical properties of baryon
  resonances in the holographic QCD}}},\ \href {\doibase 10.18910/92156} {Ph.D.
  thesis},\ \bibinfo  {school} {Osaka U.} (\bibinfo {year} {2023})\BibitemShut
  {NoStop}%
\bibitem [{\citenamefont {Bigazzi}\ and\ \citenamefont
  {Castellani}(2024)}]{Bigazzi:2023odl}%
  \BibitemOpen
  \bibfield  {author} {\bibinfo {author} {\bibfnamefont {Francesco}\
  \bibnamefont {Bigazzi}}\ and\ \bibinfo {author} {\bibfnamefont {Federico}\
  \bibnamefont {Castellani}},\ }\bibfield  {title} {\enquote {\bibinfo {title}
  {{Resonance contributions to nucleon spin structure in holographic QCD}},}\
  }\href {\doibase 10.1007/JHEP04(2024)037} {\bibfield  {journal} {\bibinfo
  {journal} {JHEP}\ }\textbf {\bibinfo {volume} {04}},\ \bibinfo {pages} {037}
  (\bibinfo {year} {2024})},\ \Eprint {http://arxiv.org/abs/2308.16833}
  {arXiv:2308.16833 [hep-ph]} \BibitemShut {NoStop}%
\bibitem [{\citenamefont {Hechenberger}\ \emph {et~al.}(2023)\citenamefont
  {Hechenberger}, \citenamefont {Leutgeb},\ and\ \citenamefont
  {Rebhan}}]{Hechenberger:2023ljn}%
  \BibitemOpen
  \bibfield  {author} {\bibinfo {author} {\bibfnamefont {Florian}\ \bibnamefont
  {Hechenberger}}, \bibinfo {author} {\bibfnamefont {Josef}\ \bibnamefont
  {Leutgeb}}, \ and\ \bibinfo {author} {\bibfnamefont {Anton}\ \bibnamefont
  {Rebhan}},\ }\bibfield  {title} {\enquote {\bibinfo {title} {{Radiative meson
  and glueball decays in the Witten-Sakai-Sugimoto model}},}\ }\href {\doibase
  10.1103/PhysRevD.107.114020} {\bibfield  {journal} {\bibinfo  {journal}
  {Phys. Rev. D}\ }\textbf {\bibinfo {volume} {107}},\ \bibinfo {pages}
  {114020} (\bibinfo {year} {2023})},\ \Eprint
  {http://arxiv.org/abs/2302.13379} {arXiv:2302.13379 [hep-ph]} \BibitemShut
  {NoStop}%
\bibitem [{\citenamefont {Ramalho}\ and\ \citenamefont
  {Pe\~na}(2024)}]{Ramalho:2023hqd}%
  \BibitemOpen
  \bibfield  {author} {\bibinfo {author} {\bibfnamefont {G.}~\bibnamefont
  {Ramalho}}\ and\ \bibinfo {author} {\bibfnamefont {M.~T.}\ \bibnamefont
  {Pe\~na}},\ }\bibfield  {title} {\enquote {\bibinfo {title} {{Electromagnetic
  transition form factors of baryon resonances}},}\ }\href {\doibase
  10.1016/j.ppnp.2024.104097} {\bibfield  {journal} {\bibinfo  {journal} {Prog.
  Part. Nucl. Phys.}\ }\textbf {\bibinfo {volume} {136}},\ \bibinfo {pages}
  {104097} (\bibinfo {year} {2024})},\ \Eprint
  {http://arxiv.org/abs/2306.13900} {arXiv:2306.13900 [hep-ph]} \BibitemShut
  {NoStop}%
\bibitem [{\citenamefont {Castellani}(2024)}]{Castellani:2024dxr}%
  \BibitemOpen
  \bibfield  {author} {\bibinfo {author} {\bibfnamefont {Federico}\
  \bibnamefont {Castellani}},\ }\bibfield  {title} {\enquote {\bibinfo {title}
  {{Nucleon electric and magnetic polarizabilities in Holographic QCD}},}\
  }\href@noop {} {\  (\bibinfo {year} {2024})},\ \Eprint
  {http://arxiv.org/abs/2402.07553} {arXiv:2402.07553 [hep-ph]} \BibitemShut
  {NoStop}%
\bibitem [{\citenamefont {Hechenberger}\ \emph {et~al.}(2024)\citenamefont
  {Hechenberger}, \citenamefont {Leutgeb},\ and\ \citenamefont
  {Rebhan}}]{Hechenberger:2024piy}%
  \BibitemOpen
  \bibfield  {author} {\bibinfo {author} {\bibfnamefont {Florian}\ \bibnamefont
  {Hechenberger}}, \bibinfo {author} {\bibfnamefont {Josef}\ \bibnamefont
  {Leutgeb}}, \ and\ \bibinfo {author} {\bibfnamefont {Anton}\ \bibnamefont
  {Rebhan}},\ }\bibfield  {title} {\enquote {\bibinfo {title} {{Spin-1
  glueballs in the Witten-Sakai-Sugimoto model}},}\ }\href {\doibase
  10.1103/PhysRevD.109.074014} {\bibfield  {journal} {\bibinfo  {journal}
  {Phys. Rev. D}\ }\textbf {\bibinfo {volume} {109}},\ \bibinfo {pages}
  {074014} (\bibinfo {year} {2024})},\ \Eprint
  {http://arxiv.org/abs/2401.17986} {arXiv:2401.17986 [hep-ph]} \BibitemShut
  {NoStop}%
\bibitem [{\citenamefont {Fujita}\ \emph {et~al.}(2022)\citenamefont {Fujita},
  \citenamefont {Hatta}, \citenamefont {Sugimoto},\ and\ \citenamefont
  {Ueda}}]{Fujita:2022jus}%
  \BibitemOpen
  \bibfield  {author} {\bibinfo {author} {\bibfnamefont {Mitsutoshi}\
  \bibnamefont {Fujita}}, \bibinfo {author} {\bibfnamefont {Yoshitaka}\
  \bibnamefont {Hatta}}, \bibinfo {author} {\bibfnamefont {Shigeki}\
  \bibnamefont {Sugimoto}}, \ and\ \bibinfo {author} {\bibfnamefont {Takahiro}\
  \bibnamefont {Ueda}},\ }\bibfield  {title} {\enquote {\bibinfo {title}
  {{Nucleon D-term in holographic quantum chromodynamics}},}\ }\href {\doibase
  10.1093/ptep/ptac110} {\bibfield  {journal} {\bibinfo  {journal} {PTEP}\
  }\textbf {\bibinfo {volume} {2022}},\ \bibinfo {pages} {093B06} (\bibinfo
  {year} {2022})},\ \Eprint {http://arxiv.org/abs/2206.06578} {arXiv:2206.06578
  [hep-th]} \BibitemShut {NoStop}%
\bibitem [{\citenamefont {Abidin}\ and\ \citenamefont
  {Carlson}(2008{\natexlab{a}})}]{Abidin:2008hn}%
  \BibitemOpen
  \bibfield  {author} {\bibinfo {author} {\bibfnamefont {Zainul}\ \bibnamefont
  {Abidin}}\ and\ \bibinfo {author} {\bibfnamefont {Carl~E.}\ \bibnamefont
  {Carlson}},\ }\bibfield  {title} {\enquote {\bibinfo {title} {{Gravitational
  Form Factors in the Axial Sector from an AdS/QCD Model}},}\ }\href {\doibase
  10.1103/PhysRevD.77.115021} {\bibfield  {journal} {\bibinfo  {journal} {Phys.
  Rev. D}\ }\textbf {\bibinfo {volume} {77}},\ \bibinfo {pages} {115021}
  (\bibinfo {year} {2008}{\natexlab{a}})},\ \Eprint
  {http://arxiv.org/abs/0804.0214} {arXiv:0804.0214 [hep-ph]} \BibitemShut
  {NoStop}%
\bibitem [{\citenamefont {Abidin}\ and\ \citenamefont
  {Carlson}(2008{\natexlab{b}})}]{Abidin:2008ku}%
  \BibitemOpen
  \bibfield  {author} {\bibinfo {author} {\bibfnamefont {Zainul}\ \bibnamefont
  {Abidin}}\ and\ \bibinfo {author} {\bibfnamefont {Carl~E.}\ \bibnamefont
  {Carlson}},\ }\bibfield  {title} {\enquote {\bibinfo {title} {{Gravitational
  form factors of vector mesons in an AdS/QCD model}},}\ }\href {\doibase
  10.1103/PhysRevD.77.095007} {\bibfield  {journal} {\bibinfo  {journal} {Phys.
  Rev. D}\ }\textbf {\bibinfo {volume} {77}},\ \bibinfo {pages} {095007}
  (\bibinfo {year} {2008}{\natexlab{b}})},\ \Eprint
  {http://arxiv.org/abs/0801.3839} {arXiv:0801.3839 [hep-ph]} \BibitemShut
  {NoStop}%
\bibitem [{\citenamefont {Abidin}\ and\ \citenamefont
  {Carlson}(2009)}]{Abidin:2009hr}%
  \BibitemOpen
  \bibfield  {author} {\bibinfo {author} {\bibfnamefont {Zainul}\ \bibnamefont
  {Abidin}}\ and\ \bibinfo {author} {\bibfnamefont {Carl~E.}\ \bibnamefont
  {Carlson}},\ }\bibfield  {title} {\enquote {\bibinfo {title} {{Nucleon
  electromagnetic and gravitational form factors from holography}},}\ }\href
  {\doibase 10.1103/PhysRevD.79.115003} {\bibfield  {journal} {\bibinfo
  {journal} {Phys. Rev. D}\ }\textbf {\bibinfo {volume} {79}},\ \bibinfo
  {pages} {115003} (\bibinfo {year} {2009})},\ \Eprint
  {http://arxiv.org/abs/0903.4818} {arXiv:0903.4818 [hep-ph]} \BibitemShut
  {NoStop}%
\bibitem [{\citenamefont {Mamo}\ and\ \citenamefont
  {Zahed}(2020)}]{Mamo:2019mka}%
  \BibitemOpen
  \bibfield  {author} {\bibinfo {author} {\bibfnamefont {Kiminad~A.}\
  \bibnamefont {Mamo}}\ and\ \bibinfo {author} {\bibfnamefont {Ismail}\
  \bibnamefont {Zahed}},\ }\bibfield  {title} {\enquote {\bibinfo {title}
  {{Diffractive photoproduction of $J/\psi$ and $\Upsilon$ using holographic
  QCD: gravitational form factors and GPD of gluons in the proton}},}\ }\href
  {\doibase 10.1103/PhysRevD.101.086003} {\bibfield  {journal} {\bibinfo
  {journal} {Phys. Rev. D}\ }\textbf {\bibinfo {volume} {101}},\ \bibinfo
  {pages} {086003} (\bibinfo {year} {2020})},\ \Eprint
  {http://arxiv.org/abs/1910.04707} {arXiv:1910.04707 [hep-ph]} \BibitemShut
  {NoStop}%
\bibitem [{\citenamefont {Chakrabarti}\ \emph {et~al.}(2020)\citenamefont
  {Chakrabarti}, \citenamefont {Mondal}, \citenamefont {Mukherjee},
  \citenamefont {Nair},\ and\ \citenamefont {Zhao}}]{Chakrabarti:2020kdc}%
  \BibitemOpen
  \bibfield  {author} {\bibinfo {author} {\bibfnamefont {Dipankar}\
  \bibnamefont {Chakrabarti}}, \bibinfo {author} {\bibfnamefont {Chandan}\
  \bibnamefont {Mondal}}, \bibinfo {author} {\bibfnamefont {Asmita}\
  \bibnamefont {Mukherjee}}, \bibinfo {author} {\bibfnamefont {Sreeraj}\
  \bibnamefont {Nair}}, \ and\ \bibinfo {author} {\bibfnamefont {Xingbo}\
  \bibnamefont {Zhao}},\ }\bibfield  {title} {\enquote {\bibinfo {title}
  {{Gravitational form factors and mechanical properties of proton in a
  light-front quark-diquark model}},}\ }\href {\doibase
  10.1103/PhysRevD.102.113011} {\bibfield  {journal} {\bibinfo  {journal}
  {Phys. Rev. D}\ }\textbf {\bibinfo {volume} {102}},\ \bibinfo {pages}
  {113011} (\bibinfo {year} {2020})},\ \Eprint
  {http://arxiv.org/abs/2010.04215} {arXiv:2010.04215 [hep-ph]} \BibitemShut
  {NoStop}%
\bibitem [{\citenamefont {Mamo}\ and\ \citenamefont
  {Zahed}(2021)}]{Mamo:2021krl}%
  \BibitemOpen
  \bibfield  {author} {\bibinfo {author} {\bibfnamefont {Kiminad~A.}\
  \bibnamefont {Mamo}}\ and\ \bibinfo {author} {\bibfnamefont {Ismail}\
  \bibnamefont {Zahed}},\ }\bibfield  {title} {\enquote {\bibinfo {title}
  {{Nucleon mass radii and distribution: Holographic QCD, Lattice QCD and GlueX
  data}},}\ }\href {\doibase 10.1103/PhysRevD.103.094010} {\bibfield  {journal}
  {\bibinfo  {journal} {Phys. Rev. D}\ }\textbf {\bibinfo {volume} {103}},\
  \bibinfo {pages} {094010} (\bibinfo {year} {2021})},\ \Eprint
  {http://arxiv.org/abs/2103.03186} {arXiv:2103.03186 [hep-ph]} \BibitemShut
  {NoStop}%
\bibitem [{\citenamefont {Mamo}\ and\ \citenamefont
  {Zahed}(2022)}]{Mamo:2022eui}%
  \BibitemOpen
  \bibfield  {author} {\bibinfo {author} {\bibfnamefont {Kiminad~A.}\
  \bibnamefont {Mamo}}\ and\ \bibinfo {author} {\bibfnamefont {Ismail}\
  \bibnamefont {Zahed}},\ }\bibfield  {title} {\enquote {\bibinfo {title}
  {{J/\ensuremath{\psi} near threshold in holographic QCD: A and D
  gravitational form factors}},}\ }\href {\doibase 10.1103/PhysRevD.106.086004}
  {\bibfield  {journal} {\bibinfo  {journal} {Phys. Rev. D}\ }\textbf {\bibinfo
  {volume} {106}},\ \bibinfo {pages} {086004} (\bibinfo {year} {2022})},\
  \Eprint {http://arxiv.org/abs/2204.08857} {arXiv:2204.08857 [hep-ph]}
  \BibitemShut {NoStop}%
\bibitem [{\citenamefont {Allahverdiyeva}\ and\ \citenamefont
  {Mamedov}(2023)}]{Allahverdiyeva:2023fhn}%
  \BibitemOpen
  \bibfield  {author} {\bibinfo {author} {\bibfnamefont {Minaya}\ \bibnamefont
  {Allahverdiyeva}}\ and\ \bibinfo {author} {\bibfnamefont {Shahin}\
  \bibnamefont {Mamedov}},\ }\bibfield  {title} {\enquote {\bibinfo {title}
  {{Vector meson gravitational form factors and generalized parton
  distributions at finite temperature within the soft-wall AdS/QCD model}},}\
  }\href {\doibase 10.1140/epjc/s10052-023-11607-7} {\bibfield  {journal}
  {\bibinfo  {journal} {Eur. Phys. J. C}\ }\textbf {\bibinfo {volume} {83}},\
  \bibinfo {pages} {447} (\bibinfo {year} {2023})},\ \Eprint
  {http://arxiv.org/abs/2302.03383} {arXiv:2302.03383 [hep-ph]} \BibitemShut
  {NoStop}%
\bibitem [{\citenamefont {Polyakov}\ and\ \citenamefont
  {Weiss}(1999)}]{Polyakov:1999gs}%
  \BibitemOpen
  \bibfield  {author} {\bibinfo {author} {\bibfnamefont {Maxim~V.}\
  \bibnamefont {Polyakov}}\ and\ \bibinfo {author} {\bibfnamefont
  {C.}~\bibnamefont {Weiss}},\ }\bibfield  {title} {\enquote {\bibinfo {title}
  {{Skewed and double distributions in pion and nucleon}},}\ }\href {\doibase
  10.1103/PhysRevD.60.114017} {\bibfield  {journal} {\bibinfo  {journal} {Phys.
  Rev. D}\ }\textbf {\bibinfo {volume} {60}},\ \bibinfo {pages} {114017}
  (\bibinfo {year} {1999})},\ \Eprint {http://arxiv.org/abs/hep-ph/9902451}
  {arXiv:hep-ph/9902451} \BibitemShut {NoStop}%
\bibitem [{\citenamefont {Broniowski}\ and\ \citenamefont
  {Ruiz~Arriola}(2008)}]{Broniowski:2008hx}%
  \BibitemOpen
  \bibfield  {author} {\bibinfo {author} {\bibfnamefont {Wojciech}\
  \bibnamefont {Broniowski}}\ and\ \bibinfo {author} {\bibfnamefont {Enrique}\
  \bibnamefont {Ruiz~Arriola}},\ }\bibfield  {title} {\enquote {\bibinfo
  {title} {{Gravitational and higher-order form factors of the pion in chiral
  quark models}},}\ }\href {\doibase 10.1103/PhysRevD.78.094011} {\bibfield
  {journal} {\bibinfo  {journal} {Phys. Rev. D}\ }\textbf {\bibinfo {volume}
  {78}},\ \bibinfo {pages} {094011} (\bibinfo {year} {2008})},\ \Eprint
  {http://arxiv.org/abs/0809.1744} {arXiv:0809.1744 [hep-ph]} \BibitemShut
  {NoStop}%
\bibitem [{\citenamefont {Frederico}\ \emph {et~al.}(2009)\citenamefont
  {Frederico}, \citenamefont {Pace}, \citenamefont {Pasquini},\ and\
  \citenamefont {Salme}}]{Frederico:2009fk}%
  \BibitemOpen
  \bibfield  {author} {\bibinfo {author} {\bibfnamefont {T.}~\bibnamefont
  {Frederico}}, \bibinfo {author} {\bibfnamefont {E.}~\bibnamefont {Pace}},
  \bibinfo {author} {\bibfnamefont {B.}~\bibnamefont {Pasquini}}, \ and\
  \bibinfo {author} {\bibfnamefont {G.}~\bibnamefont {Salme}},\ }\bibfield
  {title} {\enquote {\bibinfo {title} {{Pion Generalized Parton Distributions
  with covariant and Light-front constituent quark models}},}\ }\href {\doibase
  10.1103/PhysRevD.80.054021} {\bibfield  {journal} {\bibinfo  {journal} {Phys.
  Rev. D}\ }\textbf {\bibinfo {volume} {80}},\ \bibinfo {pages} {054021}
  (\bibinfo {year} {2009})},\ \Eprint {http://arxiv.org/abs/0907.5566}
  {arXiv:0907.5566 [hep-ph]} \BibitemShut {NoStop}%
\bibitem [{\citenamefont {Masjuan}\ \emph {et~al.}(2013)\citenamefont
  {Masjuan}, \citenamefont {Ruiz~Arriola},\ and\ \citenamefont
  {Broniowski}}]{Masjuan:2012sk}%
  \BibitemOpen
  \bibfield  {author} {\bibinfo {author} {\bibfnamefont {Pere}\ \bibnamefont
  {Masjuan}}, \bibinfo {author} {\bibfnamefont {Enrique}\ \bibnamefont
  {Ruiz~Arriola}}, \ and\ \bibinfo {author} {\bibfnamefont {Wojciech}\
  \bibnamefont {Broniowski}},\ }\bibfield  {title} {\enquote {\bibinfo {title}
  {{Meson dominance of hadron form factors and large-$N_c$ phenomenology}},}\
  }\href {\doibase 10.1103/PhysRevD.87.014005} {\bibfield  {journal} {\bibinfo
  {journal} {Phys. Rev. D}\ }\textbf {\bibinfo {volume} {87}},\ \bibinfo
  {pages} {014005} (\bibinfo {year} {2013})},\ \Eprint
  {http://arxiv.org/abs/1210.0760} {arXiv:1210.0760 [hep-ph]} \BibitemShut
  {NoStop}%
\bibitem [{\citenamefont {Son}\ and\ \citenamefont {Kim}(2014)}]{Son:2014sna}%
  \BibitemOpen
  \bibfield  {author} {\bibinfo {author} {\bibfnamefont {Hyeon-Dong}\
  \bibnamefont {Son}}\ and\ \bibinfo {author} {\bibfnamefont {Hyun-Chul}\
  \bibnamefont {Kim}},\ }\bibfield  {title} {\enquote {\bibinfo {title}
  {{Stability of the pion and the pattern of chiral symmetry breaking}},}\
  }\href {\doibase 10.1103/PhysRevD.90.111901} {\bibfield  {journal} {\bibinfo
  {journal} {Phys. Rev. D}\ }\textbf {\bibinfo {volume} {90}},\ \bibinfo
  {pages} {111901} (\bibinfo {year} {2014})},\ \Eprint
  {http://arxiv.org/abs/1410.1420} {arXiv:1410.1420 [hep-ph]} \BibitemShut
  {NoStop}%
\bibitem [{\citenamefont {Fanelli}\ \emph {et~al.}(2016)\citenamefont
  {Fanelli}, \citenamefont {Pace}, \citenamefont {Romanelli}, \citenamefont
  {Salme},\ and\ \citenamefont {Salmistraro}}]{Fanelli:2016aqc}%
  \BibitemOpen
  \bibfield  {author} {\bibinfo {author} {\bibfnamefont {Cristiano}\
  \bibnamefont {Fanelli}}, \bibinfo {author} {\bibfnamefont {Emanuele}\
  \bibnamefont {Pace}}, \bibinfo {author} {\bibfnamefont {Giovanni}\
  \bibnamefont {Romanelli}}, \bibinfo {author} {\bibfnamefont {Giovanni}\
  \bibnamefont {Salme}}, \ and\ \bibinfo {author} {\bibfnamefont {Marco}\
  \bibnamefont {Salmistraro}},\ }\bibfield  {title} {\enquote {\bibinfo {title}
  {{Pion Generalized Parton Distributions within a fully covariant constituent
  quark model}},}\ }\href {\doibase 10.1140/epjc/s10052-016-4101-1} {\bibfield
  {journal} {\bibinfo  {journal} {Eur. Phys. J. C}\ }\textbf {\bibinfo {volume}
  {76}},\ \bibinfo {pages} {253} (\bibinfo {year} {2016})},\ \Eprint
  {http://arxiv.org/abs/1603.04598} {arXiv:1603.04598 [hep-ph]} \BibitemShut
  {NoStop}%
\bibitem [{\citenamefont {Hudson}\ and\ \citenamefont
  {Schweitzer}(2017)}]{Hudson:2017xug}%
  \BibitemOpen
  \bibfield  {author} {\bibinfo {author} {\bibfnamefont {Jonathan}\
  \bibnamefont {Hudson}}\ and\ \bibinfo {author} {\bibfnamefont {Peter}\
  \bibnamefont {Schweitzer}},\ }\bibfield  {title} {\enquote {\bibinfo {title}
  {{D term and the structure of pointlike and composed spin-0 particles}},}\
  }\href {\doibase 10.1103/PhysRevD.96.114013} {\bibfield  {journal} {\bibinfo
  {journal} {Phys. Rev. D}\ }\textbf {\bibinfo {volume} {96}},\ \bibinfo
  {pages} {114013} (\bibinfo {year} {2017})},\ \Eprint
  {http://arxiv.org/abs/1712.05316} {arXiv:1712.05316 [hep-ph]} \BibitemShut
  {NoStop}%
\bibitem [{\citenamefont {Freese}\ \emph {et~al.}(2019)\citenamefont {Freese},
  \citenamefont {Freese}, \citenamefont {Clo\"et},\ and\ \citenamefont
  {Clo\"et}}]{Freese:2019bhb}%
  \BibitemOpen
  \bibfield  {author} {\bibinfo {author} {\bibfnamefont {Adam}\ \bibnamefont
  {Freese}}, \bibinfo {author} {\bibfnamefont {Adam}\ \bibnamefont {Freese}},
  \bibinfo {author} {\bibfnamefont {Ian~C.}\ \bibnamefont {Clo\"et}}, \ and\
  \bibinfo {author} {\bibfnamefont {Ian~C.}\ \bibnamefont {Clo\"et}},\
  }\bibfield  {title} {\enquote {\bibinfo {title} {{Gravitational form factors
  of light mesons}},}\ }\href {\doibase 10.1103/PhysRevC.100.015201} {\bibfield
   {journal} {\bibinfo  {journal} {Phys. Rev. C}\ }\textbf {\bibinfo {volume}
  {100}},\ \bibinfo {pages} {015201} (\bibinfo {year} {2019})},\ \bibinfo
  {note} {[Erratum: Phys.Rev.C 105, 059901 (2022)]},\ \Eprint
  {http://arxiv.org/abs/1903.09222} {arXiv:1903.09222 [nucl-th]} \BibitemShut
  {NoStop}%
\bibitem [{\citenamefont {Krutov}\ and\ \citenamefont
  {Troitsky}(2021)}]{Krutov:2020ewr}%
  \BibitemOpen
  \bibfield  {author} {\bibinfo {author} {\bibfnamefont {A.~F.}\ \bibnamefont
  {Krutov}}\ and\ \bibinfo {author} {\bibfnamefont {V.~E.}\ \bibnamefont
  {Troitsky}},\ }\bibfield  {title} {\enquote {\bibinfo {title} {{Pion
  gravitational form factors in a relativistic theory of composite
  particles}},}\ }\href {\doibase 10.1103/PhysRevD.103.014029} {\bibfield
  {journal} {\bibinfo  {journal} {Phys. Rev. D}\ }\textbf {\bibinfo {volume}
  {103}},\ \bibinfo {pages} {014029} (\bibinfo {year} {2021})},\ \Eprint
  {http://arxiv.org/abs/2010.11640} {arXiv:2010.11640 [hep-ph]} \BibitemShut
  {NoStop}%
\bibitem [{\citenamefont {Shuryak}\ and\ \citenamefont
  {Zahed}(2021)}]{Shuryak:2020ktq}%
  \BibitemOpen
  \bibfield  {author} {\bibinfo {author} {\bibfnamefont {Edward}\ \bibnamefont
  {Shuryak}}\ and\ \bibinfo {author} {\bibfnamefont {Ismail}\ \bibnamefont
  {Zahed}},\ }\bibfield  {title} {\enquote {\bibinfo {title} {{Nonperturbative
  quark-antiquark interactions in mesonic form factors}},}\ }\href {\doibase
  10.1103/PhysRevD.103.054028} {\bibfield  {journal} {\bibinfo  {journal}
  {Phys. Rev. D}\ }\textbf {\bibinfo {volume} {103}},\ \bibinfo {pages}
  {054028} (\bibinfo {year} {2021})},\ \Eprint
  {http://arxiv.org/abs/2008.06169} {arXiv:2008.06169 [hep-ph]} \BibitemShut
  {NoStop}%
\bibitem [{\citenamefont {de~T\'eramond}\ \emph {et~al.}(2021)\citenamefont
  {de~T\'eramond}, \citenamefont {Dosch}, \citenamefont {Liu}, \citenamefont
  {Sufian}, \citenamefont {Brodsky},\ and\ \citenamefont
  {Deur}}]{deTeramond:2021lxc}%
  \BibitemOpen
  \bibfield  {author} {\bibinfo {author} {\bibfnamefont {Guy~F.}\ \bibnamefont
  {de~T\'eramond}}, \bibinfo {author} {\bibfnamefont {H.~G.}\ \bibnamefont
  {Dosch}}, \bibinfo {author} {\bibfnamefont {Tianbo}\ \bibnamefont {Liu}},
  \bibinfo {author} {\bibfnamefont {Raza~Sabbir}\ \bibnamefont {Sufian}},
  \bibinfo {author} {\bibfnamefont {Stanley~J.}\ \bibnamefont {Brodsky}}, \
  and\ \bibinfo {author} {\bibfnamefont {Alexandre}\ \bibnamefont {Deur}}
  (\bibinfo {collaboration} {HLFHS}),\ }\bibfield  {title} {\enquote {\bibinfo
  {title} {{Gluon matter distribution in the proton and pion from extended
  holographic light-front QCD}},}\ }\href {\doibase
  10.1103/PhysRevD.104.114005} {\bibfield  {journal} {\bibinfo  {journal}
  {Phys. Rev. D}\ }\textbf {\bibinfo {volume} {104}},\ \bibinfo {pages}
  {114005} (\bibinfo {year} {2021})},\ \Eprint
  {http://arxiv.org/abs/2107.01231} {arXiv:2107.01231 [hep-ph]} \BibitemShut
  {NoStop}%
\bibitem [{\citenamefont {Raya}\ \emph {et~al.}(2022)\citenamefont {Raya},
  \citenamefont {Cui}, \citenamefont {Chang}, \citenamefont {Morgado},
  \citenamefont {Roberts},\ and\ \citenamefont
  {Rodriguez-Quintero}}]{Raya:2021zrz}%
  \BibitemOpen
  \bibfield  {author} {\bibinfo {author} {\bibfnamefont {Khepani}\ \bibnamefont
  {Raya}}, \bibinfo {author} {\bibfnamefont {Zhu-Fang}\ \bibnamefont {Cui}},
  \bibinfo {author} {\bibfnamefont {Lei}\ \bibnamefont {Chang}}, \bibinfo
  {author} {\bibfnamefont {Jose-Manuel}\ \bibnamefont {Morgado}}, \bibinfo
  {author} {\bibfnamefont {Craig~D.}\ \bibnamefont {Roberts}}, \ and\ \bibinfo
  {author} {\bibfnamefont {Jose}\ \bibnamefont {Rodriguez-Quintero}},\
  }\bibfield  {title} {\enquote {\bibinfo {title} {{Revealing pion and kaon
  structure via generalised parton distributions *}},}\ }\href {\doibase
  10.1088/1674-1137/ac3071} {\bibfield  {journal} {\bibinfo  {journal} {Chin.
  Phys. C}\ }\textbf {\bibinfo {volume} {46}},\ \bibinfo {pages} {013105}
  (\bibinfo {year} {2022})},\ \Eprint {http://arxiv.org/abs/2109.11686}
  {arXiv:2109.11686 [hep-ph]} \BibitemShut {NoStop}%
\bibitem [{\citenamefont {Tong}\ \emph {et~al.}(2021)\citenamefont {Tong},
  \citenamefont {Ma},\ and\ \citenamefont {Yuan}}]{Tong:2021ctu}%
  \BibitemOpen
  \bibfield  {author} {\bibinfo {author} {\bibfnamefont {Xuan-Bo}\ \bibnamefont
  {Tong}}, \bibinfo {author} {\bibfnamefont {Jian-Ping}\ \bibnamefont {Ma}}, \
  and\ \bibinfo {author} {\bibfnamefont {Feng}\ \bibnamefont {Yuan}},\
  }\bibfield  {title} {\enquote {\bibinfo {title} {{Gluon gravitational form
  factors at large momentum transfer}},}\ }\href {\doibase
  10.1016/j.physletb.2021.136751} {\bibfield  {journal} {\bibinfo  {journal}
  {Phys. Lett. B}\ }\textbf {\bibinfo {volume} {823}},\ \bibinfo {pages}
  {136751} (\bibinfo {year} {2021})},\ \Eprint
  {http://arxiv.org/abs/2101.02395} {arXiv:2101.02395 [hep-ph]} \BibitemShut
  {NoStop}%
\bibitem [{\citenamefont {Krutov}\ and\ \citenamefont
  {Troitsky}(2022)}]{Krutov:2022zgg}%
  \BibitemOpen
  \bibfield  {author} {\bibinfo {author} {\bibfnamefont {A.~F.}\ \bibnamefont
  {Krutov}}\ and\ \bibinfo {author} {\bibfnamefont {V.~E.}\ \bibnamefont
  {Troitsky}},\ }\bibfield  {title} {\enquote {\bibinfo {title} {{Relativistic
  composite-particle theory of the gravitational form factors of the pion:
  Quantitative results}},}\ }\href {\doibase 10.1103/PhysRevD.106.054013}
  {\bibfield  {journal} {\bibinfo  {journal} {Phys. Rev. D}\ }\textbf {\bibinfo
  {volume} {106}},\ \bibinfo {pages} {054013} (\bibinfo {year} {2022})},\
  \Eprint {http://arxiv.org/abs/2201.04991} {arXiv:2201.04991 [hep-ph]}
  \BibitemShut {NoStop}%
\bibitem [{\citenamefont {Xu}\ \emph {et~al.}(2024)\citenamefont {Xu},
  \citenamefont {Ding}, \citenamefont {Raya}, \citenamefont {Roberts},
  \citenamefont {Rodr\'\i{}guez-Quintero},\ and\ \citenamefont
  {Schmidt}}]{Xu:2023izo}%
  \BibitemOpen
  \bibfield  {author} {\bibinfo {author} {\bibfnamefont {Yin-Zhen}\
  \bibnamefont {Xu}}, \bibinfo {author} {\bibfnamefont {Minghui}\ \bibnamefont
  {Ding}}, \bibinfo {author} {\bibfnamefont {Kh\'epani}\ \bibnamefont {Raya}},
  \bibinfo {author} {\bibfnamefont {Craig~D.}\ \bibnamefont {Roberts}},
  \bibinfo {author} {\bibfnamefont {Jos\'e}\ \bibnamefont
  {Rodr\'\i{}guez-Quintero}}, \ and\ \bibinfo {author} {\bibfnamefont
  {Sebastian~M.}\ \bibnamefont {Schmidt}},\ }\bibfield  {title} {\enquote
  {\bibinfo {title} {{Pion and kaon electromagnetic and gravitational form
  factors}},}\ }\href {\doibase 10.1140/epjc/s10052-024-12518-x} {\bibfield
  {journal} {\bibinfo  {journal} {Eur. Phys. J. C}\ }\textbf {\bibinfo {volume}
  {84}},\ \bibinfo {pages} {191} (\bibinfo {year} {2024})},\ \Eprint
  {http://arxiv.org/abs/2311.14832} {arXiv:2311.14832 [hep-ph]} \BibitemShut
  {NoStop}%
\bibitem [{\citenamefont {Li}\ and\ \citenamefont {Vary}(2024)}]{Li:2023izn}%
  \BibitemOpen
  \bibfield  {author} {\bibinfo {author} {\bibfnamefont {Yang}\ \bibnamefont
  {Li}}\ and\ \bibinfo {author} {\bibfnamefont {James~P.}\ \bibnamefont
  {Vary}},\ }\bibfield  {title} {\enquote {\bibinfo {title} {{Stress inside the
  pion in holographic light-front QCD}},}\ }\href {\doibase
  10.1103/PhysRevD.109.L051501} {\bibfield  {journal} {\bibinfo  {journal}
  {Phys. Rev. D}\ }\textbf {\bibinfo {volume} {109}},\ \bibinfo {pages}
  {L051501} (\bibinfo {year} {2024})},\ \Eprint
  {http://arxiv.org/abs/2312.02543} {arXiv:2312.02543 [hep-th]} \BibitemShut
  {NoStop}%
\bibitem [{\citenamefont {Broniowski}\ and\ \citenamefont
  {Ruiz~Arriola}(2024)}]{Broniowski:2024oyk}%
  \BibitemOpen
  \bibfield  {author} {\bibinfo {author} {\bibfnamefont {Wojciech}\
  \bibnamefont {Broniowski}}\ and\ \bibinfo {author} {\bibfnamefont {Enrique}\
  \bibnamefont {Ruiz~Arriola}},\ }\bibfield  {title} {\enquote {\bibinfo
  {title} {{Gravitational form factors of the pion and meson dominance}},}\
  }\href@noop {} {\  (\bibinfo {year} {2024})},\ \Eprint
  {http://arxiv.org/abs/2405.07815} {arXiv:2405.07815 [hep-ph]} \BibitemShut
  {NoStop}%
\bibitem [{\citenamefont {Liu}\ \emph {et~al.}(2024{\natexlab{a}})\citenamefont
  {Liu}, \citenamefont {Shuryak}, \citenamefont {Weiss},\ and\ \citenamefont
  {Zahed}}]{Liu:2024jno}%
  \BibitemOpen
  \bibfield  {author} {\bibinfo {author} {\bibfnamefont {Wei-Yang}\
  \bibnamefont {Liu}}, \bibinfo {author} {\bibfnamefont {Edward}\ \bibnamefont
  {Shuryak}}, \bibinfo {author} {\bibfnamefont {Christian}\ \bibnamefont
  {Weiss}}, \ and\ \bibinfo {author} {\bibfnamefont {Ismail}\ \bibnamefont
  {Zahed}},\ }\bibfield  {title} {\enquote {\bibinfo {title} {{Pion
  gravitational form factors in the QCD instanton vacuum I}},}\ }\href@noop {}
  {\  (\bibinfo {year} {2024}{\natexlab{a}})},\ \Eprint
  {http://arxiv.org/abs/2405.14026} {arXiv:2405.14026 [hep-ph]} \BibitemShut
  {NoStop}%
\bibitem [{\citenamefont {Liu}\ \emph {et~al.}(2024{\natexlab{b}})\citenamefont
  {Liu}, \citenamefont {Shuryak},\ and\ \citenamefont {Zahed}}]{Liu:2024vkj}%
  \BibitemOpen
  \bibfield  {author} {\bibinfo {author} {\bibfnamefont {Wei-Yang}\
  \bibnamefont {Liu}}, \bibinfo {author} {\bibfnamefont {Edward}\ \bibnamefont
  {Shuryak}}, \ and\ \bibinfo {author} {\bibfnamefont {Ismail}\ \bibnamefont
  {Zahed}},\ }\bibfield  {title} {\enquote {\bibinfo {title} {{Pion
  gravitational form factors in the QCD instanton vacuum II}},}\ }\href@noop {}
  {\  (\bibinfo {year} {2024}{\natexlab{b}})},\ \Eprint
  {http://arxiv.org/abs/2405.16269} {arXiv:2405.16269 [hep-ph]} \BibitemShut
  {NoStop}%
\bibitem [{\citenamefont {Donoghue}\ and\ \citenamefont
  {Leutwyler}(1991)}]{Donoghue:1991qv}%
  \BibitemOpen
  \bibfield  {author} {\bibinfo {author} {\bibfnamefont {John~F.}\ \bibnamefont
  {Donoghue}}\ and\ \bibinfo {author} {\bibfnamefont {H.}~\bibnamefont
  {Leutwyler}},\ }\bibfield  {title} {\enquote {\bibinfo {title} {{Energy and
  momentum in chiral theories}},}\ }\href {\doibase 10.1007/BF01560453}
  {\bibfield  {journal} {\bibinfo  {journal} {Z. Phys. C}\ }\textbf {\bibinfo
  {volume} {52}},\ \bibinfo {pages} {343--351} (\bibinfo {year}
  {1991})}\BibitemShut {NoStop}%
\bibitem [{\citenamefont {Burkert}(2018)}]{Burkert:2018nvj}%
  \BibitemOpen
  \bibfield  {author} {\bibinfo {author} {\bibfnamefont {Volker~D.}\
  \bibnamefont {Burkert}},\ }\bibfield  {title} {\enquote {\bibinfo {title}
  {{Jefferson Lab at 12 GeV: The Science Program}},}\ }\href {\doibase
  10.1146/annurev-nucl-101917-021129} {\bibfield  {journal} {\bibinfo
  {journal} {Ann. Rev. Nucl. Part. Sci.}\ }\textbf {\bibinfo {volume} {68}},\
  \bibinfo {pages} {405--428} (\bibinfo {year} {2018})}\BibitemShut {NoStop}%
\bibitem [{\citenamefont {Abdul~Khalek}\ \emph {et~al.}(2022)\citenamefont
  {Abdul~Khalek} \emph {et~al.}}]{AbdulKhalek:2021gbh}%
  \BibitemOpen
  \bibfield  {author} {\bibinfo {author} {\bibfnamefont {R.}~\bibnamefont
  {Abdul~Khalek}} \emph {et~al.},\ }\bibfield  {title} {\enquote {\bibinfo
  {title} {{Science Requirements and Detector Concepts for the Electron-Ion
  Collider}: {EIC Yellow Report}},}\ }\href {\doibase
  10.1016/j.nuclphysa.2022.122447} {\bibfield  {journal} {\bibinfo  {journal}
  {Nucl. Phys. A}\ }\textbf {\bibinfo {volume} {1026}},\ \bibinfo {pages}
  {122447} (\bibinfo {year} {2022})},\ \Eprint
  {http://arxiv.org/abs/2103.05419} {arXiv:2103.05419 [physics.ins-det]}
  \BibitemShut {NoStop}%
\bibitem [{\citenamefont {Anderle}\ \emph {et~al.}(2021)\citenamefont {Anderle}
  \emph {et~al.}}]{Anderle:2021wcy}%
  \BibitemOpen
  \bibfield  {author} {\bibinfo {author} {\bibfnamefont {Daniele~P.}\
  \bibnamefont {Anderle}} \emph {et~al.},\ }\bibfield  {title} {\enquote
  {\bibinfo {title} {{Electron-ion collider in China}},}\ }\href {\doibase
  10.1007/s11467-021-1062-0} {\bibfield  {journal} {\bibinfo  {journal} {Front.
  Phys. (Beijing)}\ }\textbf {\bibinfo {volume} {16}},\ \bibinfo {pages}
  {64701} (\bibinfo {year} {2021})},\ \Eprint {http://arxiv.org/abs/2102.09222}
  {arXiv:2102.09222 [nucl-ex]} \BibitemShut {NoStop}%
\bibitem [{\citenamefont {Georges}\ \emph {et~al.}(2022)\citenamefont {Georges}
  \emph {et~al.}}]{JeffersonLabHallA:2022pnx}%
  \BibitemOpen
  \bibfield  {author} {\bibinfo {author} {\bibfnamefont {F.}~\bibnamefont
  {Georges}} \emph {et~al.} (\bibinfo {collaboration} {Jefferson Lab Hall A}),\
  }\bibfield  {title} {\enquote {\bibinfo {title} {{Deeply Virtual Compton
  Scattering Cross Section at High Bjorken xB}},}\ }\href {\doibase
  10.1103/PhysRevLett.128.252002} {\bibfield  {journal} {\bibinfo  {journal}
  {Phys. Rev. Lett.}\ }\textbf {\bibinfo {volume} {128}},\ \bibinfo {pages}
  {252002} (\bibinfo {year} {2022})},\ \Eprint
  {http://arxiv.org/abs/2201.03714} {arXiv:2201.03714 [hep-ph]} \BibitemShut
  {NoStop}%
\bibitem [{\citenamefont {Christiaens}\ \emph {et~al.}(2023)\citenamefont
  {Christiaens} \emph {et~al.}}]{CLAS:2022syx}%
  \BibitemOpen
  \bibfield  {author} {\bibinfo {author} {\bibfnamefont {G.}~\bibnamefont
  {Christiaens}} \emph {et~al.} (\bibinfo {collaboration} {CLAS}),\ }\bibfield
  {title} {\enquote {\bibinfo {title} {{First CLAS12 Measurement of Deeply
  Virtual Compton Scattering Beam-Spin Asymmetries in the Extended Valence
  Region}},}\ }\href {\doibase 10.1103/PhysRevLett.130.211902} {\bibfield
  {journal} {\bibinfo  {journal} {Phys. Rev. Lett.}\ }\textbf {\bibinfo
  {volume} {130}},\ \bibinfo {pages} {211902} (\bibinfo {year} {2023})},\
  \Eprint {http://arxiv.org/abs/2211.11274} {arXiv:2211.11274 [hep-ex]}
  \BibitemShut {NoStop}%
\bibitem [{\citenamefont {Kobzarev}\ and\ \citenamefont
  {Okun}(1962)}]{Kobzarev:1962wt}%
  \BibitemOpen
  \bibfield  {author} {\bibinfo {author} {\bibfnamefont {I.~Yu.}\ \bibnamefont
  {Kobzarev}}\ and\ \bibinfo {author} {\bibfnamefont {L.~B.}\ \bibnamefont
  {Okun}},\ }\bibfield  {title} {\enquote {\bibinfo {title} {{GRAVITATIONAL
  INTERACTION OF FERMIONS}},}\ }\href@noop {} {\bibfield  {journal} {\bibinfo
  {journal} {Zh. Eksp. Teor. Fiz.}\ }\textbf {\bibinfo {volume} {43}},\
  \bibinfo {pages} {1904--1909} (\bibinfo {year} {1962})}\BibitemShut {NoStop}%
\bibitem [{\citenamefont {Pagels}(1966)}]{Pagels:1966zza}%
  \BibitemOpen
  \bibfield  {author} {\bibinfo {author} {\bibfnamefont {Heinz}\ \bibnamefont
  {Pagels}},\ }\bibfield  {title} {\enquote {\bibinfo {title} {{Energy-Momentum
  Structure Form Factors of Particles}},}\ }\href {\doibase
  10.1103/PhysRev.144.1250} {\bibfield  {journal} {\bibinfo  {journal} {Phys.
  Rev.}\ }\textbf {\bibinfo {volume} {144}},\ \bibinfo {pages} {1250--1260}
  (\bibinfo {year} {1966})}\BibitemShut {NoStop}%
\bibitem [{\citenamefont {Witten}(1998{\natexlab{a}})}]{Witten:1998zw}%
  \BibitemOpen
  \bibfield  {author} {\bibinfo {author} {\bibfnamefont {Edward}\ \bibnamefont
  {Witten}},\ }\bibfield  {title} {\enquote {\bibinfo {title} {{Anti-de Sitter
  space, thermal phase transition, and confinement in gauge theories}},}\
  }\href {\doibase 10.4310/ATMP.1998.v2.n3.a3} {\bibfield  {journal} {\bibinfo
  {journal} {Adv. Theor. Math. Phys.}\ }\textbf {\bibinfo {volume} {2}},\
  \bibinfo {pages} {505--532} (\bibinfo {year} {1998}{\natexlab{a}})},\ \Eprint
  {http://arxiv.org/abs/hep-th/9803131} {arXiv:hep-th/9803131} \BibitemShut
  {NoStop}%
\bibitem [{\citenamefont {de~Haro}\ \emph {et~al.}(2001)\citenamefont
  {de~Haro}, \citenamefont {Solodukhin},\ and\ \citenamefont
  {Skenderis}}]{deHaro:2000vlm}%
  \BibitemOpen
  \bibfield  {author} {\bibinfo {author} {\bibfnamefont {Sebastian}\
  \bibnamefont {de~Haro}}, \bibinfo {author} {\bibfnamefont {Sergey~N.}\
  \bibnamefont {Solodukhin}}, \ and\ \bibinfo {author} {\bibfnamefont {Kostas}\
  \bibnamefont {Skenderis}},\ }\bibfield  {title} {\enquote {\bibinfo {title}
  {{Holographic reconstruction of space-time and renormalization in the AdS /
  CFT correspondence}},}\ }\href {\doibase 10.1007/s002200100381} {\bibfield
  {journal} {\bibinfo  {journal} {Commun. Math. Phys.}\ }\textbf {\bibinfo
  {volume} {217}},\ \bibinfo {pages} {595--622} (\bibinfo {year} {2001})},\
  \Eprint {http://arxiv.org/abs/hep-th/0002230} {arXiv:hep-th/0002230}
  \BibitemShut {NoStop}%
\bibitem [{\citenamefont {Kanitscheider}\ \emph {et~al.}(2008)\citenamefont
  {Kanitscheider}, \citenamefont {Skenderis},\ and\ \citenamefont
  {Taylor}}]{Kanitscheider:2008kd}%
  \BibitemOpen
  \bibfield  {author} {\bibinfo {author} {\bibfnamefont {Ingmar}\ \bibnamefont
  {Kanitscheider}}, \bibinfo {author} {\bibfnamefont {Kostas}\ \bibnamefont
  {Skenderis}}, \ and\ \bibinfo {author} {\bibfnamefont {Marika}\ \bibnamefont
  {Taylor}},\ }\bibfield  {title} {\enquote {\bibinfo {title} {{Precision
  holography for non-conformal branes}},}\ }\href {\doibase
  10.1088/1126-6708/2008/09/094} {\bibfield  {journal} {\bibinfo  {journal}
  {JHEP}\ }\textbf {\bibinfo {volume} {09}},\ \bibinfo {pages} {094} (\bibinfo
  {year} {2008})},\ \Eprint {http://arxiv.org/abs/0807.3324} {arXiv:0807.3324
  [hep-th]} \BibitemShut {NoStop}%
\bibitem [{\citenamefont {Gubser}\ \emph {et~al.}(1998)\citenamefont {Gubser},
  \citenamefont {Klebanov},\ and\ \citenamefont {Polyakov}}]{Gubser:1998bc}%
  \BibitemOpen
  \bibfield  {author} {\bibinfo {author} {\bibfnamefont {S.~S.}\ \bibnamefont
  {Gubser}}, \bibinfo {author} {\bibfnamefont {Igor~R.}\ \bibnamefont
  {Klebanov}}, \ and\ \bibinfo {author} {\bibfnamefont {Alexander~M.}\
  \bibnamefont {Polyakov}},\ }\bibfield  {title} {\enquote {\bibinfo {title}
  {{Gauge theory correlators from noncritical string theory}},}\ }\href
  {\doibase 10.1016/S0370-2693(98)00377-3} {\bibfield  {journal} {\bibinfo
  {journal} {Phys. Lett. B}\ }\textbf {\bibinfo {volume} {428}},\ \bibinfo
  {pages} {105--114} (\bibinfo {year} {1998})},\ \Eprint
  {http://arxiv.org/abs/hep-th/9802109} {arXiv:hep-th/9802109} \BibitemShut
  {NoStop}%
\bibitem [{\citenamefont {Witten}(1998{\natexlab{b}})}]{Witten:1998qj}%
  \BibitemOpen
  \bibfield  {author} {\bibinfo {author} {\bibfnamefont {Edward}\ \bibnamefont
  {Witten}},\ }\bibfield  {title} {\enquote {\bibinfo {title} {{Anti-de Sitter
  space and holography}},}\ }\href {\doibase 10.4310/ATMP.1998.v2.n2.a2}
  {\bibfield  {journal} {\bibinfo  {journal} {Adv. Theor. Math. Phys.}\
  }\textbf {\bibinfo {volume} {2}},\ \bibinfo {pages} {253--291} (\bibinfo
  {year} {1998}{\natexlab{b}})},\ \Eprint {http://arxiv.org/abs/hep-th/9802150}
  {arXiv:hep-th/9802150} \BibitemShut {NoStop}%
\bibitem [{\citenamefont {Casero}\ \emph {et~al.}(2007)\citenamefont {Casero},
  \citenamefont {Kiritsis},\ and\ \citenamefont {Paredes}}]{Casero:2007ae}%
  \BibitemOpen
  \bibfield  {author} {\bibinfo {author} {\bibfnamefont {Roberto}\ \bibnamefont
  {Casero}}, \bibinfo {author} {\bibfnamefont {Elias}\ \bibnamefont
  {Kiritsis}}, \ and\ \bibinfo {author} {\bibfnamefont {Angel}\ \bibnamefont
  {Paredes}},\ }\bibfield  {title} {\enquote {\bibinfo {title} {{Chiral
  symmetry breaking as open string tachyon condensation}},}\ }\href {\doibase
  10.1016/j.nuclphysb.2007.07.009} {\bibfield  {journal} {\bibinfo  {journal}
  {Nucl. Phys. B}\ }\textbf {\bibinfo {volume} {787}},\ \bibinfo {pages}
  {98--134} (\bibinfo {year} {2007})},\ \Eprint
  {http://arxiv.org/abs/hep-th/0702155} {arXiv:hep-th/0702155} \BibitemShut
  {NoStop}%
\bibitem [{\citenamefont {Hashimoto}\ \emph {et~al.}(2007)\citenamefont
  {Hashimoto}, \citenamefont {Hirayama},\ and\ \citenamefont
  {Miwa}}]{Hashimoto:2007fa}%
  \BibitemOpen
  \bibfield  {author} {\bibinfo {author} {\bibfnamefont {Koji}\ \bibnamefont
  {Hashimoto}}, \bibinfo {author} {\bibfnamefont {Takayuki}\ \bibnamefont
  {Hirayama}}, \ and\ \bibinfo {author} {\bibfnamefont {Akitsugu}\ \bibnamefont
  {Miwa}},\ }\bibfield  {title} {\enquote {\bibinfo {title} {{Holographic QCD
  and pion mass}},}\ }\href {\doibase 10.1088/1126-6708/2007/06/020} {\bibfield
   {journal} {\bibinfo  {journal} {JHEP}\ }\textbf {\bibinfo {volume} {06}},\
  \bibinfo {pages} {020} (\bibinfo {year} {2007})},\ \Eprint
  {http://arxiv.org/abs/hep-th/0703024} {arXiv:hep-th/0703024} \BibitemShut
  {NoStop}%
\bibitem [{\citenamefont {Bergman}\ \emph {et~al.}(2007)\citenamefont
  {Bergman}, \citenamefont {Seki},\ and\ \citenamefont
  {Sonnenschein}}]{Bergman:2007pm}%
  \BibitemOpen
  \bibfield  {author} {\bibinfo {author} {\bibfnamefont {Oren}\ \bibnamefont
  {Bergman}}, \bibinfo {author} {\bibfnamefont {Shigenori}\ \bibnamefont
  {Seki}}, \ and\ \bibinfo {author} {\bibfnamefont {Jacob}\ \bibnamefont
  {Sonnenschein}},\ }\bibfield  {title} {\enquote {\bibinfo {title} {{Quark
  mass and condensate in HQCD}},}\ }\href {\doibase
  10.1088/1126-6708/2007/12/037} {\bibfield  {journal} {\bibinfo  {journal}
  {JHEP}\ }\textbf {\bibinfo {volume} {12}},\ \bibinfo {pages} {037} (\bibinfo
  {year} {2007})},\ \Eprint {http://arxiv.org/abs/0708.2839} {arXiv:0708.2839
  [hep-th]} \BibitemShut {NoStop}%
\bibitem [{\citenamefont {Dhar}\ and\ \citenamefont
  {Nag}(2008{\natexlab{a}})}]{Dhar:2007bz}%
  \BibitemOpen
  \bibfield  {author} {\bibinfo {author} {\bibfnamefont {Avinash}\ \bibnamefont
  {Dhar}}\ and\ \bibinfo {author} {\bibfnamefont {Partha}\ \bibnamefont
  {Nag}},\ }\bibfield  {title} {\enquote {\bibinfo {title} {{Sakai-Sugimoto
  model, Tachyon Condensation and Chiral symmetry Breaking}},}\ }\href
  {\doibase 10.1088/1126-6708/2008/01/055} {\bibfield  {journal} {\bibinfo
  {journal} {JHEP}\ }\textbf {\bibinfo {volume} {01}},\ \bibinfo {pages} {055}
  (\bibinfo {year} {2008}{\natexlab{a}})},\ \Eprint
  {http://arxiv.org/abs/0708.3233} {arXiv:0708.3233 [hep-th]} \BibitemShut
  {NoStop}%
\bibitem [{\citenamefont {Hashimoto}\ \emph
  {et~al.}(2008{\natexlab{c}})\citenamefont {Hashimoto}, \citenamefont
  {Hirayama}, \citenamefont {Lin},\ and\ \citenamefont
  {Yee}}]{Hashimoto:2008sr}%
  \BibitemOpen
  \bibfield  {author} {\bibinfo {author} {\bibfnamefont {Koji}\ \bibnamefont
  {Hashimoto}}, \bibinfo {author} {\bibfnamefont {Takayuki}\ \bibnamefont
  {Hirayama}}, \bibinfo {author} {\bibfnamefont {Feng-Li}\ \bibnamefont {Lin}},
  \ and\ \bibinfo {author} {\bibfnamefont {Ho-Ung}\ \bibnamefont {Yee}},\
  }\bibfield  {title} {\enquote {\bibinfo {title} {{Quark Mass Deformation of
  Holographic Massless QCD}},}\ }\href {\doibase 10.1088/1126-6708/2008/07/089}
  {\bibfield  {journal} {\bibinfo  {journal} {JHEP}\ }\textbf {\bibinfo
  {volume} {07}},\ \bibinfo {pages} {089} (\bibinfo {year}
  {2008}{\natexlab{c}})},\ \Eprint {http://arxiv.org/abs/0803.4192}
  {arXiv:0803.4192 [hep-th]} \BibitemShut {NoStop}%
\bibitem [{\citenamefont {Dhar}\ and\ \citenamefont
  {Nag}(2008{\natexlab{b}})}]{Dhar:2008um}%
  \BibitemOpen
  \bibfield  {author} {\bibinfo {author} {\bibfnamefont {Avinash}\ \bibnamefont
  {Dhar}}\ and\ \bibinfo {author} {\bibfnamefont {Partha}\ \bibnamefont
  {Nag}},\ }\bibfield  {title} {\enquote {\bibinfo {title} {{Tachyon
  condensation and quark mass in modified Sakai-Sugimoto model}},}\ }\href
  {\doibase 10.1103/PhysRevD.78.066021} {\bibfield  {journal} {\bibinfo
  {journal} {Phys. Rev. D}\ }\textbf {\bibinfo {volume} {78}},\ \bibinfo
  {pages} {066021} (\bibinfo {year} {2008}{\natexlab{b}})},\ \Eprint
  {http://arxiv.org/abs/0804.4807} {arXiv:0804.4807 [hep-th]} \BibitemShut
  {NoStop}%
\bibitem [{\citenamefont {McNees}\ \emph {et~al.}(2008)\citenamefont {McNees},
  \citenamefont {Myers},\ and\ \citenamefont {Sinha}}]{McNees:2008km}%
  \BibitemOpen
  \bibfield  {author} {\bibinfo {author} {\bibfnamefont {Robert}\ \bibnamefont
  {McNees}}, \bibinfo {author} {\bibfnamefont {Robert~C.}\ \bibnamefont
  {Myers}}, \ and\ \bibinfo {author} {\bibfnamefont {Aninda}\ \bibnamefont
  {Sinha}},\ }\bibfield  {title} {\enquote {\bibinfo {title} {{On quark masses
  in holographic QCD}},}\ }\href {\doibase 10.1088/1126-6708/2008/11/056}
  {\bibfield  {journal} {\bibinfo  {journal} {JHEP}\ }\textbf {\bibinfo
  {volume} {11}},\ \bibinfo {pages} {056} (\bibinfo {year} {2008})},\ \Eprint
  {http://arxiv.org/abs/0807.5127} {arXiv:0807.5127 [hep-th]} \BibitemShut
  {NoStop}%
\end{thebibliography}%

\end{document}